\documentclass[aps,prb,twocolumn,groupedaddress,showpacs]{revtex4}
\usepackage{epsfig}
\usepackage[dvipsnames,usenames]{color}
\usepackage[normalem]{ulem}

\emergencystretch=\maxdimen
\begin{document}

\title{Determinant Quantum Monte Carlo Study of $d$-wave pairing
in the Plaquette Hubbard Hamiltonian}

\author{T.~Ying$^{1}$, R.~Mondaini$^{2,3}$, X.D.~Sun$^1$,
T.~Paiva$^{4}$, R.M.~Fye$^{5}$ and R.T.~Scalettar$^3$}

\affiliation{$^1$Department of Physics, Harbin Institute of Technology, Harbin
150001, China}

\affiliation{$^2$Physics Department, The Pennsylvania State University, 104
Davey Laboratory, University Park, Pennsylvania 16802, USA}

\affiliation{$^3$Physics Department, University of California, Davis,
California 95616, USA}

\affiliation{$^4$Instituto de Fisica, Universidade Federal do Rio de
Janeiro Cx.P. 68.528, 21941-972 Rio de Janeiro RJ, Brazil}

\affiliation{$^5$ Sandia National Laboratories, New
Mexico, P.O.~Box 5800
Albuquerque, NM 87185}

\begin{abstract}
Determinant Quantum Monte Carlo (DQMC) is used to determine the pairing and
magnetic response for a Hubbard model built up from four-site clusters -
a two-dimensional square lattice consisting of elemental 2x2 plaquettes
with hopping $t$ and on-site repulsion $U$ coupled by an inter-plaquette
hopping $t' \leq t$.  Superconductivity in this geometry has previously
been studied by a variety of analytic and numeric methods, with
differing conclusions concerning whether the pairing correlations and
transition temperature are raised near half-filling
by the inhomogeneous hopping or not.  For $U/t=4$,
DQMC indicates an optimal $t'/t \approx 0.4$ at which the pairing vertex
is most attractive.
The optimal $t'/t$ increases with $U/t$.
We then contrast our results for this plaquette model with a Hamiltonian
which instead involves a regular pattern of site energies whose
large site energy limit is the three band CuO$_2$ model; we
show that there the inhomogeneity rapidly, and monotonically,
suppresses pairing.
\end{abstract}

\pacs{
71.10.Fd, 
02.70.Uu  
}
\maketitle

\section{Introduction}

One of the earliest numerical indications of the possibility that an
on-site electron-electron interaction $U$ might play a role in superconducting materials was the observation of a negative ``binding
energy" in exact diagonalization studies of the Hubbard Hamiltonian on
2x2 clusters.  In this geometry, the ground state
energy of two holes doped together into a half-filled system was shown
to be lower than if the two holes were on separate clusters:
\begin{equation}
\Delta_p = E_0(2) + E_0(0) - 2\, E_0(1) < 0
\label{eq:bindenergy}
\end{equation}
Here
$E_0(n)$ is the ground state energy of $n$ holes.  The observation that
the $n=2$ and $n=0$ ground states have $s$- and $d$-wave symmetry,
respectively, and hence are connected by a $d$-wave pair creation
operator, suggested the possible relevance of models involving such 2x2
clusters with cuprate superconductors.\cite{scalapino96}  Pair binding
was also studied on larger Hubbard clusters,\cite{riera89,dagotto90} and
on other geometries, e.g., on one dimensional chains of varying
length,\cite{fye90} with three electronic
bands,\cite{hirsch88a,balseiro88,hirsch89a} models with intersite
interactions,\cite{callaway90} and the strong coupling $t$-$J$
limit.\cite{riera89,kaxiras88,bonca89,hasegawa89}

Following these small cluster studies, a considerable amount of analytic and
numeric attention has been focused on the ``plaquette Hubbard model"
which consists of a periodic array of 2x2 plaquettes with hopping $t$
and repulsion $U$ connected by a weaker hybridization $t'$.  It
was suggested that the
plaquettes act as centers of attraction, which then
drive superconductivity in
the extended lattice.  This picture provides a `local'
counterpart to theories of pairing which focus qualitatively on the
exchange of magnetic fluctuations.  Perhaps unsurprisingly, the presence
of inhomogeneous hoppings introduces phases to the Mott insulator,
antiferromagnetic, and $d$-wave superconductor typically discussed in
the uniform $t=t'$ case.  Specifically, the quantum numbers
and symmetries of the 2x2 plaquette can evolve into a wide variety of
ground states when $t'$ is made nonzero.\cite{yao07}
An additional diagonal hopping
can also change the ground state of the 2x2 plaquette
building block\cite{yao10} and induce types of crystalline
insulators. The effects of both chemical potential and hopping
disorder on pair binding have been examined,\cite{smith13}
and were shown to be less damaging
to superconductivity when there is a plaquette structure
compared to the uniform case.

A key conceptual question concerns the existence of an `optimal
inhomogeneity'.\cite{arrigoni04,martin05,kivelson06}
As pointed out by Tsai and Kivelson,\cite{tsai06}
pairing which exists at very weak $t'$ is expected to exhibit
a critical temperature $T_c$ which increases as $t'$ grows.
If it were the case that $T_c$ is small or zero in the homogeneous
model $t'=t$, this necessarily implies a maximal $T_c$ at
an intermediate value $0 < t'/t < 1$.
Early work relevant to this issue looked at pair binding energies
when two plaquettes were linked in different geometries \cite{fye92}.
For the cubic (fully connected) configuration, a maximum
binding was found for $t'/t \approx 0.3$ at $U/t=4$ and
for $t'/t \approx 0.5$ when $U/t=8$.
Exact diagonalization of 4x4 clusters\cite{tsai08} indicated that
the overall maximum occurs at $t'/t \approx 0.5$ and $U/t \approx 8$.
Additional evidence for an optimal inhomogeneity in the
plaquette Hubbard model is provided by a
contractor-renormalization (CORE) study\cite{baruch10}
where the pair binding energy was found to be maximized in the
range $0.5 < t'/t < 0.7$ and $5 < U/t < 8$.

In related work,
the density matrix renormalization group method has been used to
study a collection of 2x2 plaquettes connected to form a
two leg ladder.\cite{karakonstantakis11}
It was found that, close to half-filling, $U/t \approx 6$
and $t'/t \approx 0.6$ gives the optimal pair binding energy.
Although there can be no finite temperature transition in such
one-dimensional ladder geometries,
an interchain mean field theory
suggests that the critical temperature again exhibits an
`optimal degree of inhomogeneity' with a maximum occurring at
$t' < t$.

There have also been several methods which
challenge the idea of an optimal inhomogeneity at intermediate
$t'/t$.
The central result of a
Dynamical Cluster Approximation (DCA) analysis\cite{doluweera08}
was that the critical temperature $T_c$
for $d$-wave pairing is maximal for $t'/t=1$ for interaction
strengths $U$ of the order of the bandwidth and lattice
fillings $\rho \approx 0.9$.
That is, inhomogeneity monotonically suppresses superconductivity.
The qualitative physical picture behind this conclusion
was that inhomogeneities
reduce the magnetic contributions to the pairing
interaction.\cite{maier06,maier07a,maier07b}

Cellular Dynamical Mean Field Theory (CDMFT) is another approach with
which the plaquette Hubbard Hamiltonian has been analyzed.\cite{chakraborty11}
At weak coupling, inhomogeneity reduces the order
parameter for small to intermediate doping, but enhances it at larger
doping.  For strong coupling, inhomogeneity suppresses pairing for all
doping.  Overall, the CDMFT results seem generally consistent with those of the
DCA, namely that for inhomogeneity in the nearest-neighbor hopping such
as is present in the plaquette Hubbard model, the superconducting order
parameter does not exceed that of the uniform system.

The contrasting results between the DMRG interchain MFT, CORE, and exact
diagonalization treatments, and other cases in which optimal
inhomogeneity occurs, on one hand, and the DCA, CDMFT methods on the
other, provide the motivation for the work described in
this manuscript
- a study of the plaquette Hubbard Hamiltonian\cite{foot1} using the
Determinant Quantum Monte Carlo method.\cite{blankenbecler81,white89a}
After presenting our results,
we will discuss some of the possible origins of
the range of conclusions concerning the underlying physics of this
model.

The remainder of this paper is organized as follows: In Sec.~II we write
down the plaquette Hubbard Hamiltonian and discuss the measurements we
use to monitor $d$-wave pairing.  We also provide a brief summary of the
DQMC algorithm and its limitations.  In Sec.~III we discuss our  results at
half-filling and in the doped case.  Our central conclusion is
that an optimal degree of inhomogeneity does occur in the plaquette
Hubbard model, although the largest pairing signal appears to occur at
$t'/t \approx 0.4$ for $U/t=4$, a bit less than that reported in other work.
This optimal $t'/t$ increases with $U/t$ at half-filling.
The sign problem restricts us
to higher temperatures than those accessible in the
DCA\cite{doluweera08} and CDMFT\cite{chakraborty11} approaches.  Section
IV discusses the effect on pairing of another form of inhomogeneity in
which the site energies are varied periodically across the lattice,
and Sec.~V concerns the sign problem.  The
paper concludes with a summary of our findings.

\section{The Plaquette Hubbard Hamiltonian}

\begin{figure}[t]
\epsfig{figure=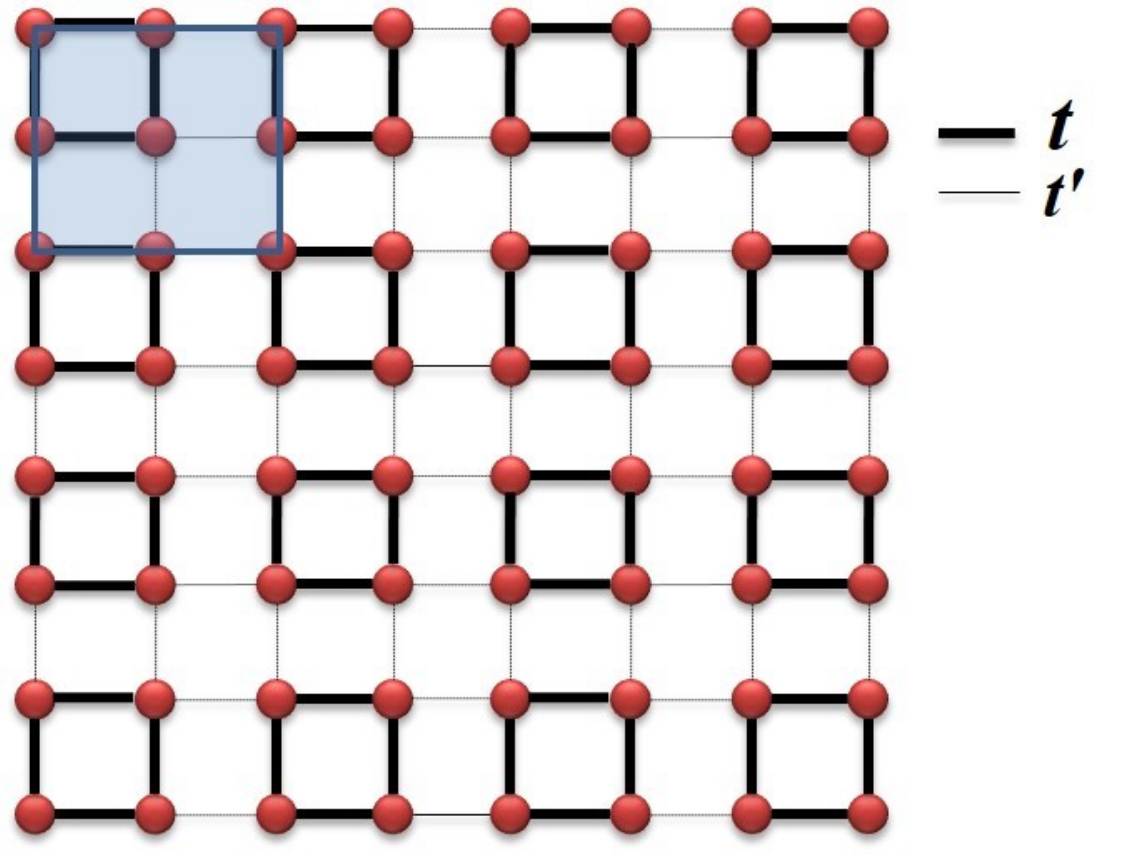,width=6.0cm,angle=-0,clip}
\caption{(Color online)
Lattice geometry for the
plaquette Hubbard model, a 2D square lattice built from
plaquettes of strong hopping $t$ connected by weaker hopping $t'$.
\label{fig:geometries}
}
\end{figure}

The plaquette Hubbard Hamiltonian is
\begin{eqnarray}
\hat H = &-&t \sum_{\langle ij\rangle \in {\cal P},\sigma} \big( \,
c_{i\sigma}^{\dagger} c_{j\sigma}^{\phantom{\dagger}}
+ c_{j\sigma}^{\dagger} c_{i\sigma}^{\phantom{\dagger}}
\, \big)
\nonumber \\
&-&t' \sum_{\langle ij\rangle \not\in {\cal P},\sigma} \big( \,
c_{i\sigma}^{\dagger} c_{j\sigma}^{\phantom{\dagger}}
+ c_{j\sigma}^{\dagger} c_{i\sigma}^{\phantom{\dagger}}
\, \big)
\label{eq:plaquetteham}
\\
&+&U \sum_i \left(n_{i\uparrow} - \frac12\right) \left(n_{i\downarrow} -
\frac12\right)
-\mu \sum_i (n_{i\uparrow} + n_{i\downarrow})
\nonumber
\end{eqnarray}
Here $c_{i\sigma}^{\dagger}\,(c_{i\sigma}^{\phantom{\dagger}})$ are the
usual creation(destruction) operators for fermions of spin $\sigma$ on
lattice site $i$.  The designations $\langle ij \rangle \in {\cal P}$
and $\langle ij \rangle \not\in {\cal P}$ in the kinetic energy terms
convey the fact that hopping $t$ between near neighbor sites $i,j$ on the same
plaquette is different from the hopping $t'$ for sites $i,j$ on
different plaquettes.  This geometry is illustrated in
Fig.~\ref{fig:geometries}.  We have written the interaction term
in particle-hole symmetric form, so that $\mu=0$ corresponds to
half-filling.  (Note that the Hubbard Hamiltonian with near-neighbor
hopping on a bipartite lattice is particle-hole symmetric for {\it any}
pattern of intersite hoppings $t_{ij}$, and hence, in particular, for
the case considered here.)

Although we have referred to $t'$ as the `interplaquette hopping',
so that $t'=0$ is the limit of independent 2x2 clusters, we note that
setting $t=0$ also results in a collection of
decoupled 2x2 $t'$ clusters.  More generally, the Hamiltonian is
invariant\cite{doluweera08,chakraborty11}
under the interchange of $t$ and $t'$.  As a consequence,
there is no need to explore the physics of $t'/t>1$.
The Hamiltonian is also invariant when the values of $t$ and $t'$ are
interchanged only on the horizontal links, or only on the vertical
links.  Our numerical approach preserves all these symmetries.

In the Determinant Quantum Monte Carlo (DQMC) algorithm
\cite{blankenbecler81,white89a},
the expectation values of observables
$\langle \hat A \rangle = {\rm Tr} \,\,  \hat A \,
{\rm exp}(-\beta \hat H) \, / \,
{\rm Tr} \,\,  {\rm exp}(-\beta \hat H)$
for fermonic Hamiltonians like Eq.~\ref{eq:plaquetteham} are evaluated by
discretizing the inverse temperature $\beta$ and
rewriting the partition function as a path integral.
Replacing the exponential of
the interaction terms in the Hamiltonian by
a coupling of quadratic fermion operators to
a Hubbard-Stratonovich field allows the fermions
to be integrated out analytically, leaving a product of fermion
determinants (one determinant for each spin species) as
the weight to sample the Hubbard-Stratonovich field.
Each operator $\hat A$ can then be measured by accumulating appropriate
combinations of Green's functions, the inverse of the matrices
whose determinants form the Boltzmann weight.
As described further below, the flexibility to alter
the order in which the Monte Carlo average is performed and
in which the Green's functions are multiplied
can be used to control which many body effects
are included in the expectation value, and hence to isolate the
pairing vertex.

The discretization of $\beta$ introduces a `Trotter error'.
We have used $\Delta \tau = 1/8$
in the work reported here\cite{hirsch85,white89a}.
In practice, unless one examines
a local quantity like the energy or double occupancy which
can be obtained to very high accuracy, the systematic Trotter
errors with this choice of $\Delta \tau$ are less than the
statistical errors in the measurements we present.

\begin{figure}[t]
\epsfig{figure=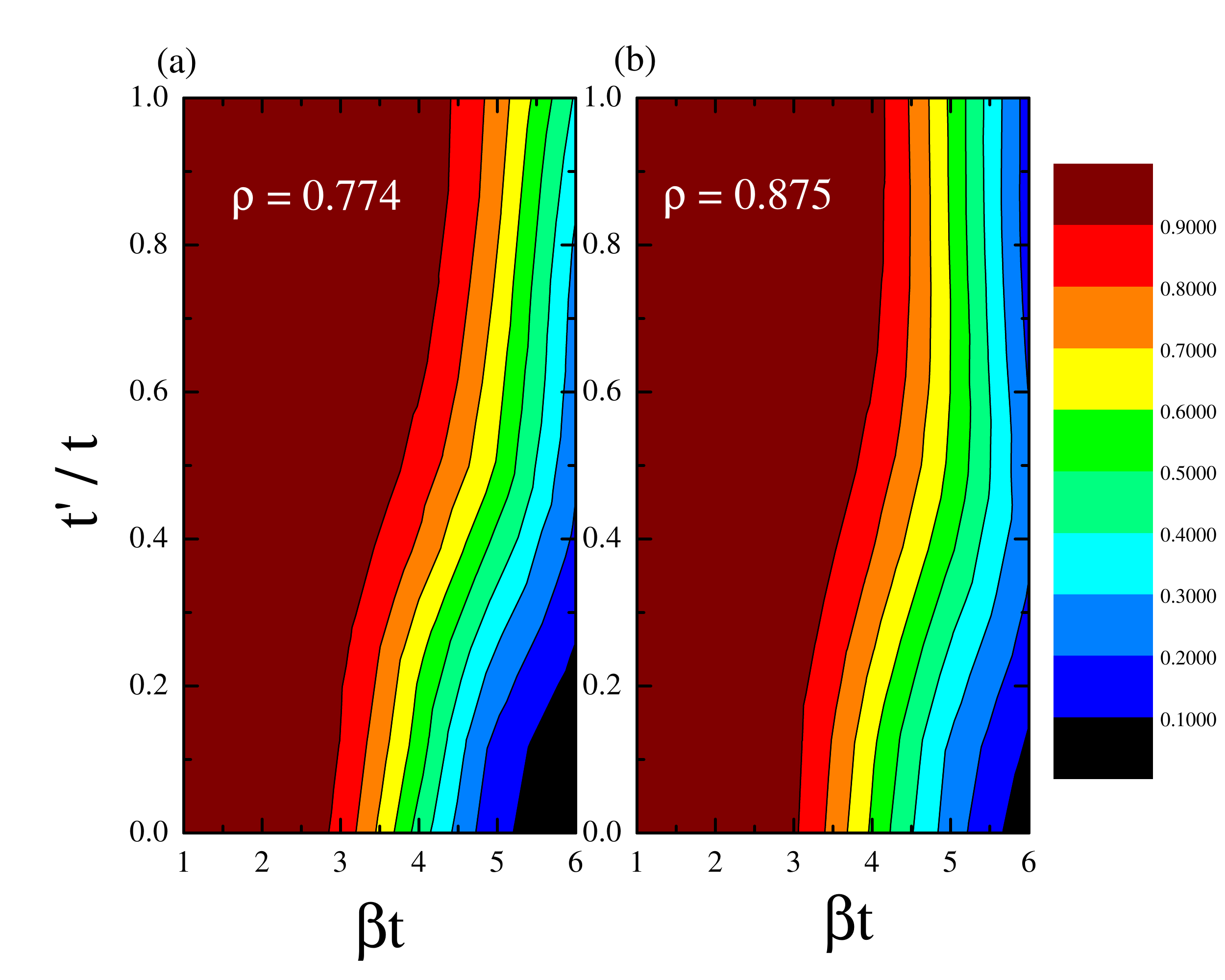,width=9.0cm,angle=-0,clip}
\caption{(Color online)
The average sign
$\langle \, {\cal S} \, \rangle $
is shown for the plaquette Hubbard model at $U/t=4$
and an 8x8 lattice for two different densities.  (Because of particle-hole
symmetry $\langle \, {\cal S} \, \rangle = 1$ at $\rho=1$.)
The sign problem becomes somewhat worse with inhomogeneity
$t' \neq t$.
Roughly speaking, it becomes difficult to generate accurate data in DQMC when
$\langle \, {\cal S} \, \rangle \lesssim 0.3$.
\label{fig:signrho0774rho0875L8} 
}
\end{figure}

The central limitation to the DQMC algorithm is the sign
problem\cite{loh90}
which arises when the product of determinants becomes negative.
This will restrict the temperatures accessible in the
study reported here, and, as a consequence, temper our
ability to make conclusive statements about the effect
of inhomogeneity in the case when the system is doped.
At half-filling, because spatial variations in the hopping
do not destroy particle-hole symmetry, there is no sign problem
and DQMC can better access the ground state for any $t'/t$.
Off half-filling data for the average sign
$\langle \, {\cal S} \, \rangle $
are given in Fig.~\ref{fig:signrho0774rho0875L8}.
$\langle \, {\cal S} \, \rangle $ is relatively weakly dependent
on $t'/t$.  The lowest accessible temperature is around $T/t \sim 1/5$ for the
entire range $0 < t'/t < 1$, although simulations become somewhat more
difficult as $t'/t$ decreases.
It is possible to get accurate data for certain quantities, like the
density, for quite small values of
$\langle \, {\cal S} \, \rangle $.
However for more complex quantities like magnetic and pair correlations
at large distances,  if reasonable accuracy
(statistical error bars less than 10\%) is desired, then
$\langle \, {\cal S} \, \rangle  \gtrsim 0.3 $ is needed.
$\langle \, {\cal S} \, \rangle $
is roughly the same for the two densities
$\rho=0.875$ and $\rho=0.774$ shown. For $\rho=0.500$,
however,
$\langle \, {\cal S} \, \rangle $ is better behaved (not shown) and reliable
averages can be obtained for temperatures as low as $T/t = 1/16$, for several
values of $t'/t$.

\begin{figure}[t]
\epsfig{figure=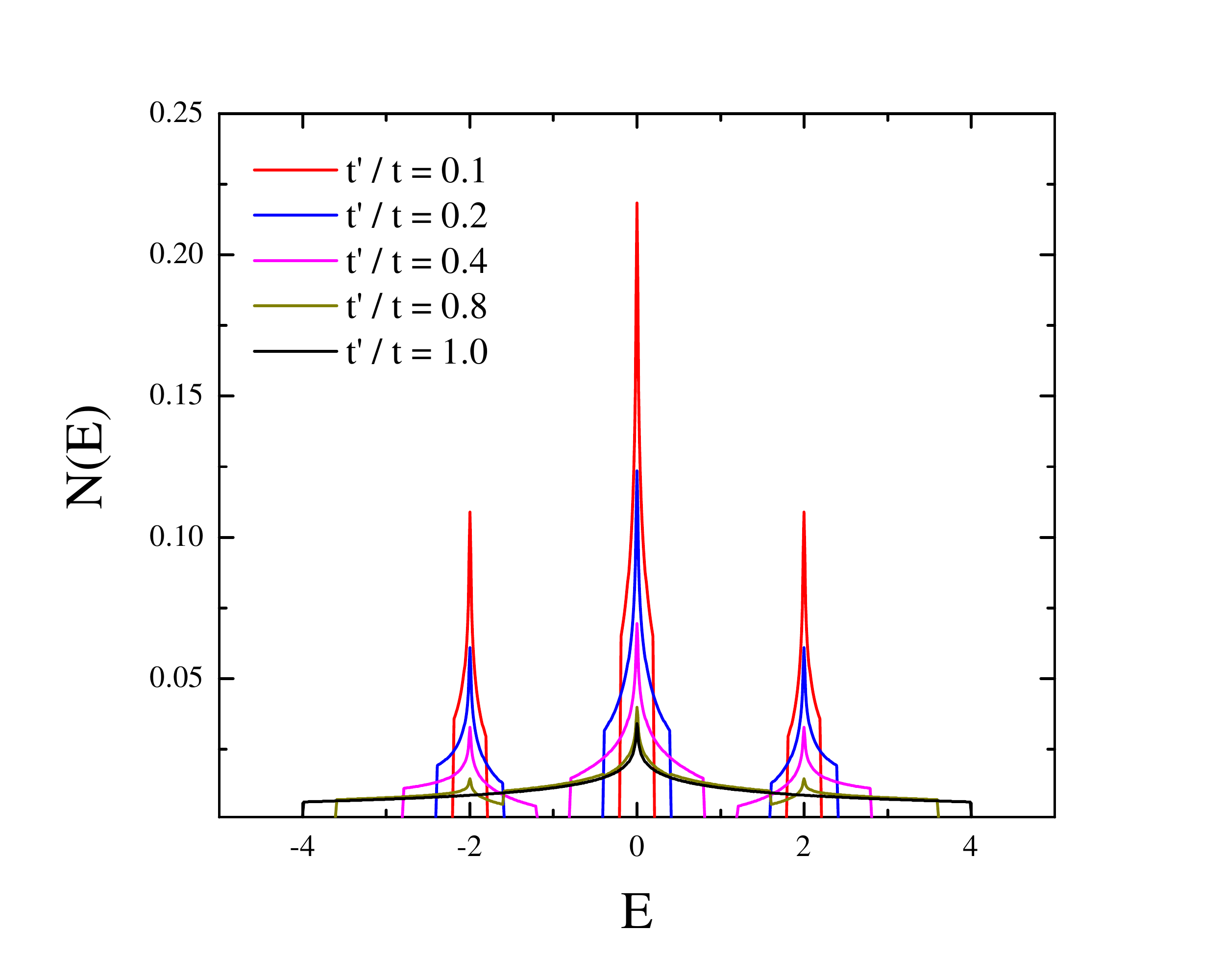,width=9.0cm,angle=-0,clip}
\caption{(Color online)
The noninteracting density of states of the uniform ($t'=t$)
2D Hubbard model
extends from $-4t$ to $+4t$ and has a van Hove
singularity at $E=0$.
In the other limit $t'/t=0$ there are four discrete (delta function) levels
at $E=-2t,0,0,+2t$.
The density of states is shown here for interpolating
ratios of $t'/t$, exhibiting the evolution between these cases.
Regardless of the relative values,
Eq.~\ref{eq:plaquetteham}
is particle hole symmetric, implying $N(E)=N(-E)$.
}
\label{fig:noninteractingdos} 
\end{figure}

The spectrum of the $U=0$ hopping Hamiltonian for
an isolated 2x2 plaquette consists of four energy levels,
$E=-2t, 0, 0, 2t$.  As $t'$ is turned on, these discrete levels
broaden until they finally merge into the 2D square lattice
density of states $N(E)$ at $t'=t$.  This evolution is shown in
Fig.~\ref{fig:noninteractingdos}.  At half-filling, where
$E_{\rm Fermi}=0$,  and  for small
dopings, $N(E_{\rm Fermi})$ is enhanced by inhomogeneity.  In principle
this might lead to a greater tendency to ordered phases, including
superconducting ones, although the possibly competing
effect of inhomogeneity on
the interaction vertex must also be considered.\cite{doluweera08}

\begin{figure}[h]
\epsfig{figure=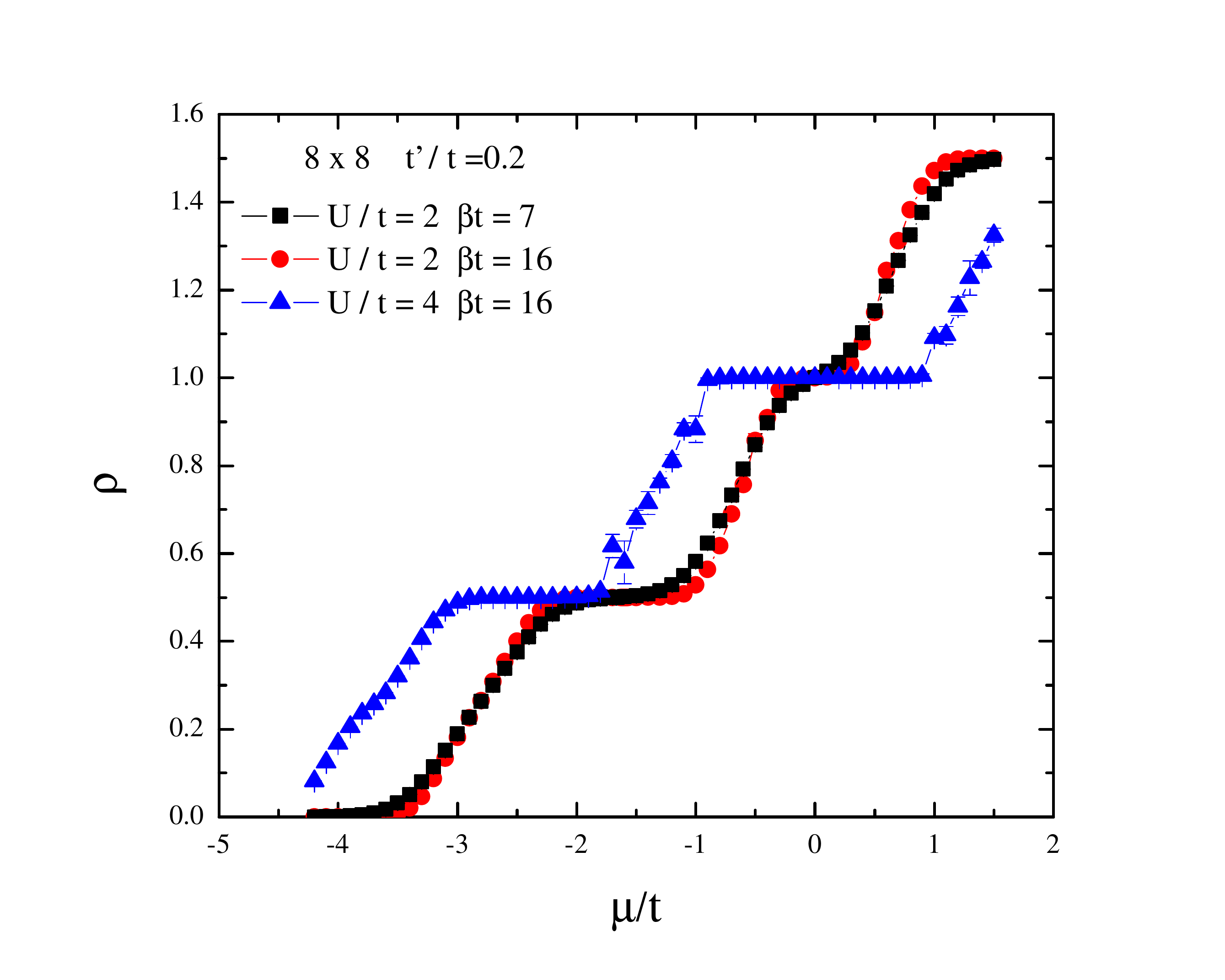,width=9.0cm,angle=-0,clip}
\caption{(Color online)
Density as a function of chemical potential for an 8x8 lattice
at $t'/t=0.2$.  The band gaps evident in the $U/t=0$ density of states
at $\rho=0.5$ and $\rho=1.5$
(Fig.~\ref{fig:noninteractingdos})  persist
at weak to intermediate coupling $U/t=2$-$4$ shown here.
(Since $\rho$ is particle-hole symmetric we focus  on $\rho\lesssim1$.)
However the interactions also drive
the formation of an insulating gap at $\rho=1.$
\label{fig:rhovsmutp02L8}  
}
\end{figure}

\begin{figure}[t]
\epsfig{figure=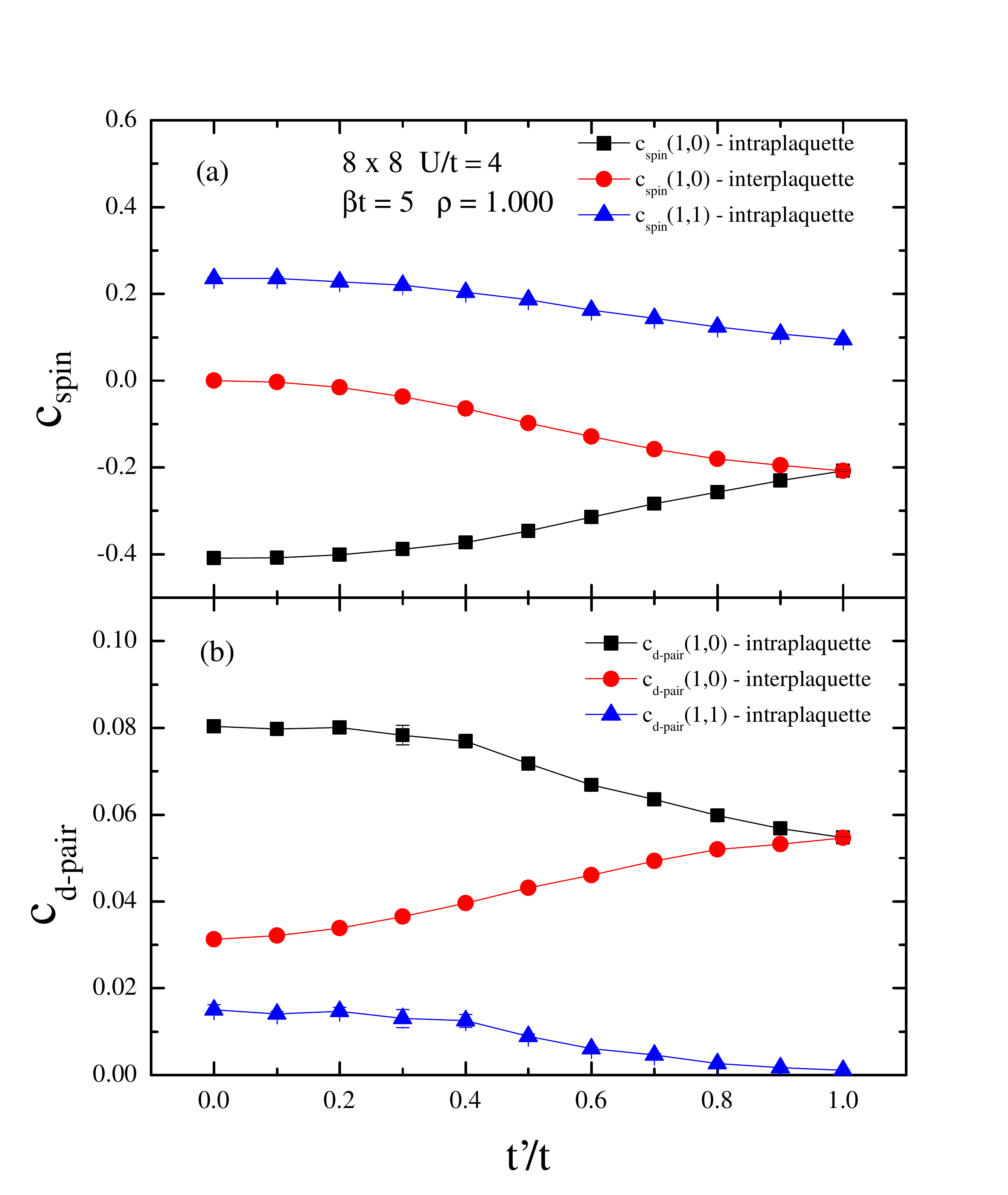,width=9.0cm,angle=-0,clip}
\caption{(Color online)
Spin (a) and charge (b) correlations along an intraplaquette bond
(black squares), interplaquette bond (red circles), and along the
diagonal of the plaquette (blue triangles).  Here $\rho=1$,
$U/t=4$,  and $\beta t =5$.  The lattice is 8x8.
\label{fig:spinandcharge} 
}
\end{figure}

For large inhomogeneity ($t'/t < 0.5$) the discrete 2x2 eigen-levels
are not sufficiently broadened by $t'$ to coalesce into a single band,
and the noninteracting
system is a band insulator at $\rho=0.5$ and $\rho=1.5$.
Figure \ref{fig:rhovsmutp02L8} shows QMC data for $\rho(\mu)$
at interaction strengths $U/t=2$ and $U/t=4$ and weakly
coupled plaquettes $t'/t=0.2$.  There is a band gap
evident at $\rho=0.5$ (and also, due to  particle-hole symmetry at
$\rho=1.5$, not shown).  Non-zero $U/t$ is also seen to
cause an insulating gap to develop at half-filling, $\rho=1$.
This is a dramatic change from the noninteracting limit, since it
represents the suppression of the large peak in $N(E)$
at $E=0$ in Fig.~\ref{fig:noninteractingdos}.
The development of this gap, even though $U/t$ is much less than
the bandwidth, is associated with the onset of long range
antiferromagnetic order, as we shall see in the next section.
Notice that reasonable data can be obtained for the density
even at $U/t=4, \beta t=16$.
This, however, is not true for more complicated spin and pair
correlations.

The equal time spin correlation function and
magnetic structure factor are given by,
\begin{eqnarray}
c_{\rm spin}({\vec r}\,)&=&
\langle \, m^{\phantom{\dagger}}_{\vec r_0 + \vec r} \,
m^{\dagger}_{\vec r_0}
\, \rangle
\hskip0.50in
m^{\dagger}_{r} = c_{\vec r\uparrow}^{\dagger}
c_{\vec r\downarrow}^{\phantom{\dagger}}
\nonumber \\
S^{+-}(q_x,q_y) &=& \frac{1}{N} \sum_{i,j} c_{\rm spin}(\vec r \,)
\,\,e^{i \vec q \cdot \vec r }
\label{eq:saf}
\end{eqnarray}
with an analogous expression for $S^{zz}(q_x,q_y)$.
In the homogeneous system it is known that at $T/t = 0$ and at half-filling
the 2D Hubbard Hamiltonian possesses long range
magnetic order.\cite{white89a,hirsch85,hirsch89,varney09}
That is, the spin-spin correlations $c_{\rm spin}(\vec r\,)$
in real space approach a
nonzero value asymptotically
as $|\, \vec r\, | \rightarrow \infty$.  On finite
sized lattices, this is established by an appropriate scaling of the
structure factor with lattice size. \cite{huse}

As with magnetic order,
a tendency to $d$-wave pairing can be examined via the asymptotic
behavior of equal time correlations,
\begin{eqnarray}
c_{d \,{\rm pair}}({\vec r}\,)&=&
\langle \Delta^{\phantom{\dagger}}_{d\, \vec r_0 + \vec r}
\, \Delta^{\dagger}_{d\,\vec r_0} \rangle
\\
\Delta^{\dagger}_{d\,\vec r} &=&
c^\dagger_{\vec r\uparrow} (c^{\dagger}_{\vec r+\hat x\downarrow}
-c^{\dagger}_{\vec r+\hat y\downarrow}
+c^{\dagger}_{\vec r-\hat x\downarrow}
-c^{\dagger}_{\vec r-\hat y\downarrow} )
\nonumber
\label{equaltime}
\end{eqnarray}
However, a more sensitive measurement, and one which makes better
contact with previous DCA work,\cite{doluweera08} is
the $d$-wave pairing susceptibility,
\begin{eqnarray}
c_{d \, \rm pair}(\vec r,\tau)&=&
\langle \Delta^{\phantom{\dagger}}_{d\,\vec r_0+\vec r}(\tau)
\Delta^\dagger_{d\,\vec r_0}(0) \rangle
\nonumber \\
\Delta^\dagger_{d\, \vec r}(\tau) &=&
e^{\tau H} \Delta^\dagger_{d\, \vec r}(0) e^{-\tau H}
\nonumber \\
P_d &=&  \sum_{\vec r} \int_0^\beta  c_{d\,{\rm pair}}(\vec r,\tau)
\,\, d\tau
\label{eq:pairsusc}
\end{eqnarray}
$P_d$ is a preferred diagnostic of superconductivity, especially if the
sign problem precludes going to low temperatures, because it allows for
a comparison between the fully dressed susceptibility and the
uncorrelated susceptibility $\overline{P}_d$, and hence
an indication of pairing even when only short range
order is present.\cite{white89b}  The technical distinction
between $P_d$ and $\overline{P}_d$ in a DQMC simulation is that when the
expectation value of the four fermion terms in Eq.~\ref{eq:pairsusc}
is evaluated, the
Green's functions obtained by the Wick contractions are first multiplied
together and then averaged to obtain $P_d$, whereas for
$\overline{P}_d$,  the Green's functions are first averaged and then
multiplied.  In $\overline{P}_d$
the effect of the interactions is only to dress the individual single
particle propagators, while $P_d$ includes all interaction
effects\cite{white89b}.

This distinction allows us to extract the interaction vertex $\Gamma_d$
from $P_d$ and $\overline{P}_d$:
\begin{eqnarray}
\Gamma_d = {1 \over P_d}
- {1 \over \overline{P}_d} \,\,.
\label{vertex}
\end{eqnarray}
If $\Gamma_d \overline P_d < 0$, the associated pairing interaction is
attractive.  More precisely, Eq.~\ref{vertex} can be re-written as,
\begin{eqnarray}
P_d = \frac{\overline P_d}{1+\Gamma_d \overline P_d}
\label{eq:stonerlike}
\end{eqnarray}
so that $\Gamma_d \overline P_d \rightarrow -1$ signals a
superconducting instability.

\section{Results}

\subsection{Half-Filling}

\begin{figure}[t]
\epsfig{figure=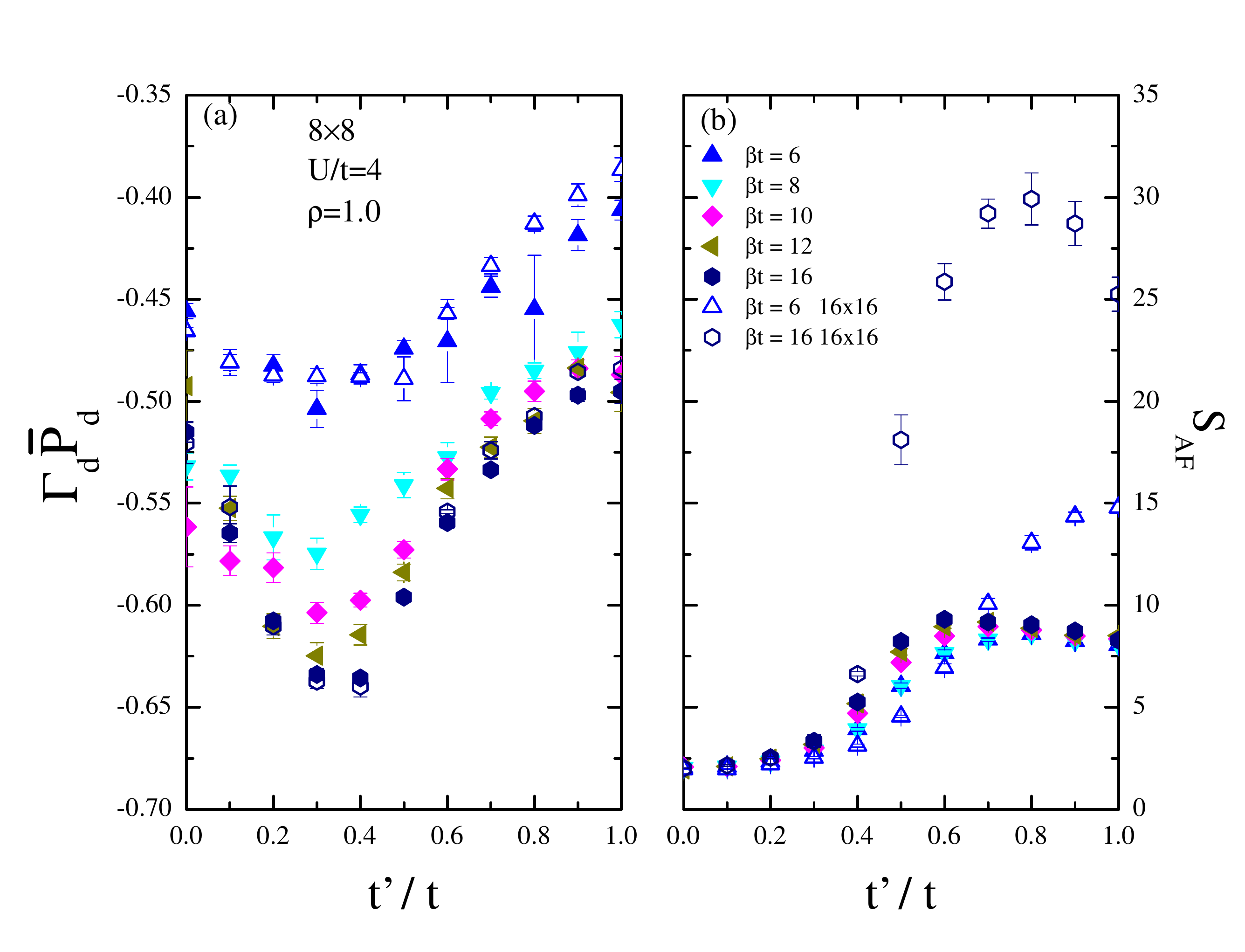,width=9.0cm,angle=-0,clip}
\caption{(Color online)
(a) Product of $d$-wave superconducting vertex $\Gamma$ and no-vertex
pairing susceptibility $\overline P_d$ as a function of
inter-plaquette hopping $t'$.
Parameters are half-filling $(\mu /t=0$) and $U/t=4$.
If $\Gamma \overline P_d \rightarrow -1$, a superconducting instability
ensues.  Pairing tendency is optimized at intermediate $t'/t \approx
0.40$, and increases as temperature is lowered.  Finite
size effects (8x8 and 16x16 lattices are compared) are minimal.
(b) Antiferromagnetic structure factor $S_{\rm SF}$.
Long range correlations (antiferromagnetic correlation length
exceeds finite size of lattice)
does not develop at $\beta t=16$ until $t'/t \gtrsim 0.6$.
\label{fig:U4mu0} 
}
\end{figure}

Our central interest is in the doped lattice, where
antiferromagnetism might potentially give way to $d$-wave pairing.
However, we begin by briefly showing results at $\rho=1$, which,
as we shall see, are not qualitatively so dissimilar to $\rho < 1$.

Due to  the spatial inhomogeneity, spin and pair correlations are not
the same on all pairs of near-neighbor (NN) links.  In
Fig.~\ref{fig:spinandcharge} (a) we show the NN spin
correlations $c_{\rm spin}(1,0)$ along an intraplaquette ($t$) bond and
along an interplaquette ($t'$) bond.  We also show the next-near
neighbor (NNN) correlation $c_{\rm spin}(1,1)$ across the internal
diagonal of a plaquette.
The NN values are negative, indicating (short-range) antiferromagnetic
order.  As expected, the interplaquette value vanishes at $t'/t=0$ and the
two NN correlations become degenerate when $t'/t=1$.  The NNN correlations
are positive, in agreement with antiferromagnetic behavior.

Figure~\ref{fig:spinandcharge}(b) shows the analogous short-range
$d$-wave pair correlations $c_{\rm pair}(1,0)$ and $c_{\rm pair}(1,1)$.
The value of $c_{\rm pair}(1,0)$ along an interplaquette ($t'$) bond does
not  vanish at $t'/t=0$ owing to the finite spatial size of the
$d$-wave operator (Eq.~\ref{eq:pairsusc}).  Short-ranged pairing correlations
change very smoothly with $t'/t$.  We will therefore turn to the
more sensitive magnetic and pairing structure factors and
susceptibilities.

\begin{figure}[t]
\epsfig{figure=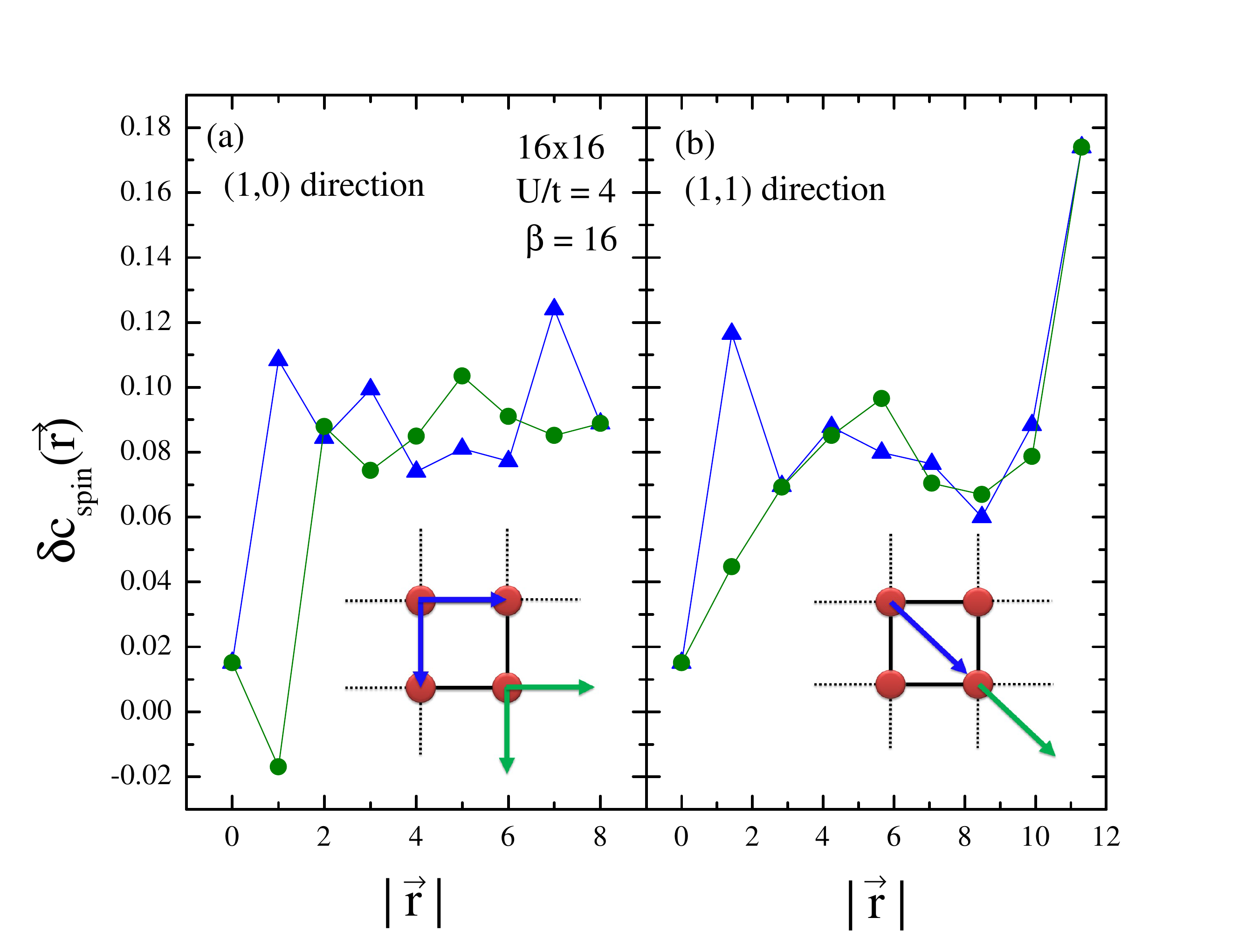,width=9.0cm,angle=-0,clip}
\caption{(Color online)
(a) Normalized difference of the spin correlation for the $t'/t =
0.8$ and the $t'/t = 1.0$ at a given position $\vec{r}$ along the equivalent NN
lines ( $(1,0)$ or $(0,1)$ ). The (green) circles are the correlations outward from a plaquette while in the (blue) triangles are the same but the correlations start in a direction inside the plaquettes, as depicted in the inset.
(b) The same as in (a) but for the $(1,1)$ direction, i.e., along the NNN links.
\label{fig:c_spin_real_space_comp} 
}
\end{figure}

The left panel of Fig.~\ref{fig:U4mu0} shows the product
$\Gamma \overline P_d$ of the pairing vertex and the uncorrelated
susceptibility.
$\Gamma \overline P_d$
becomes closest to $-1$, where a superconducting instability
would occur, at an intermediate value $t'/t \sim 0.4$.
The tendency to pairing becomes greater as $\beta t$ is increased
(lower temperature).  Finite size effects are small, with data for
8x8 and 16x16 lattices largely coinciding.

Fig.~\ref{fig:U4mu0} (b) shows the antiferromagnetic
structure factor $S_{\rm AF} \equiv S(\pi,\pi)$.  We emphasize that
$\rho=1$ is privileged from the point of the view of the DQMC algorithm,
since there is no sign problem and hence very low temperatures can be
simulated.  The large values of $S_{\rm AF}$ evident in
Fig.~\ref{fig:U4mu0}(b) arise from the development of longer ranged
correlations at the low temperatures accessible at $\rho=1$, so that the
spatial sum in Eq.~\ref{eq:saf} receives contributions from all lattice
separations.  In principle, $S_{\rm AF} \propto L^2$, the lattice
volume, but there are significant finite size corrections and a careful
scaling analysis\cite{white89a,varney09,huse} is required to establish
long range order.  Nevertheless, the growth in $S_{\rm AF}$ from $L=8$
to $L=16$ in Fig.~\ref{fig:U4mu0}(b) is certainly suggestive.  The sharp
onset at $t'/t \sim 0.6$ is similar to results reported in
[\onlinecite{baruch10}], as we shall discuss further below.  $S_{\rm
AF}$ appears to have a maximum at intermediate $t'/t$ on the 8x8
lattice, an effect which is even more pronounced in the largest size, 16x16.

\begin{figure}[t]
\epsfig{figure=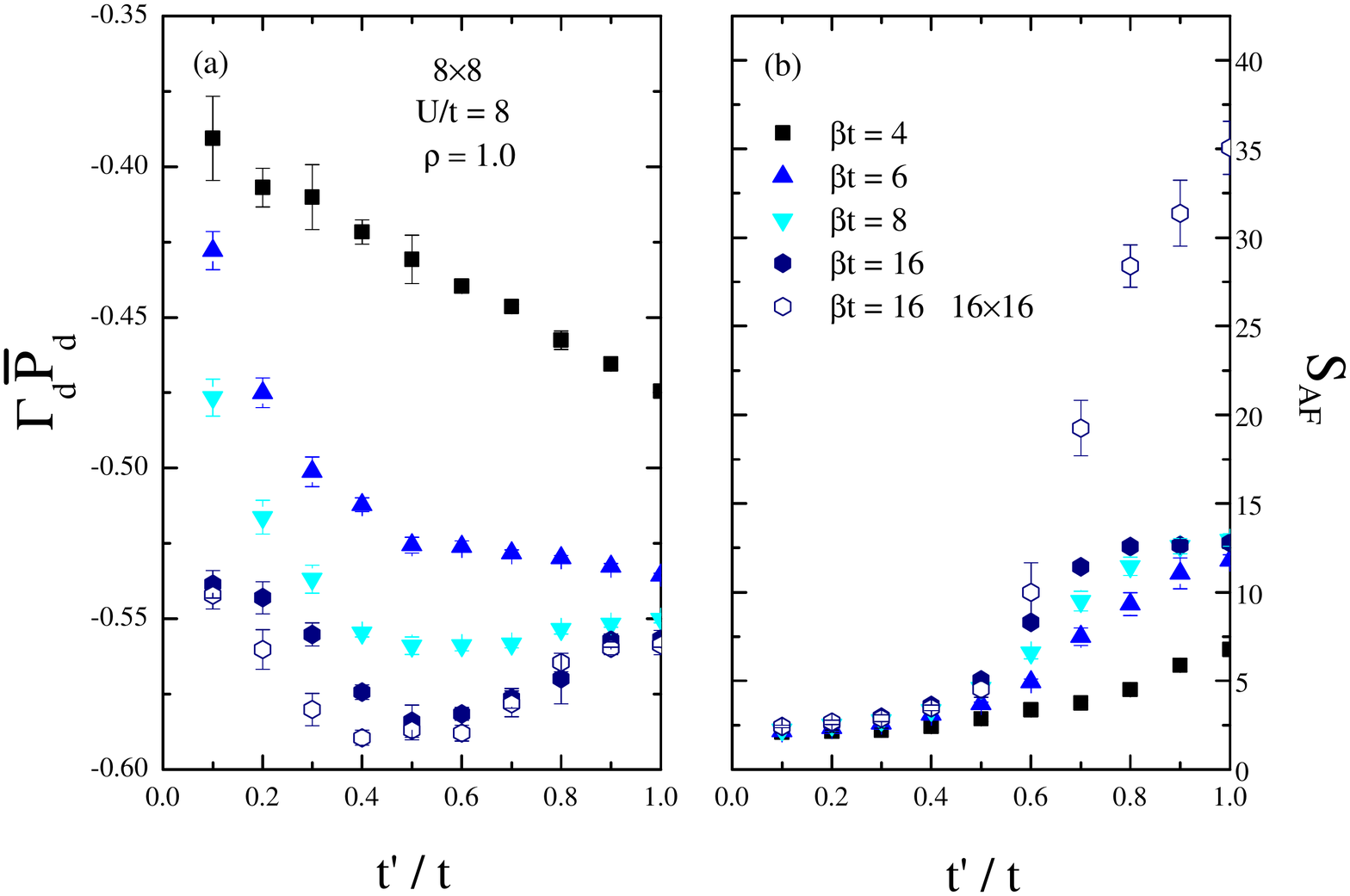,width=9.0cm,angle=-0,clip}
\caption{(Color online)
Similar to Fig.~\ref{fig:U4mu0} except $U/t=8$.  The left panel shows the
measure $\Gamma \bar P_d$ of the pairing instability as a function of
inhomogeneity $t'/t$ for different inverse temperatures, and the right
panel is the antiferromagnetic structure factor.  The lattice is
half-filled $\rho=1$.
}
\label{fig:U8mu0} 
\end{figure}

\begin{figure}[h!]
\epsfig{figure=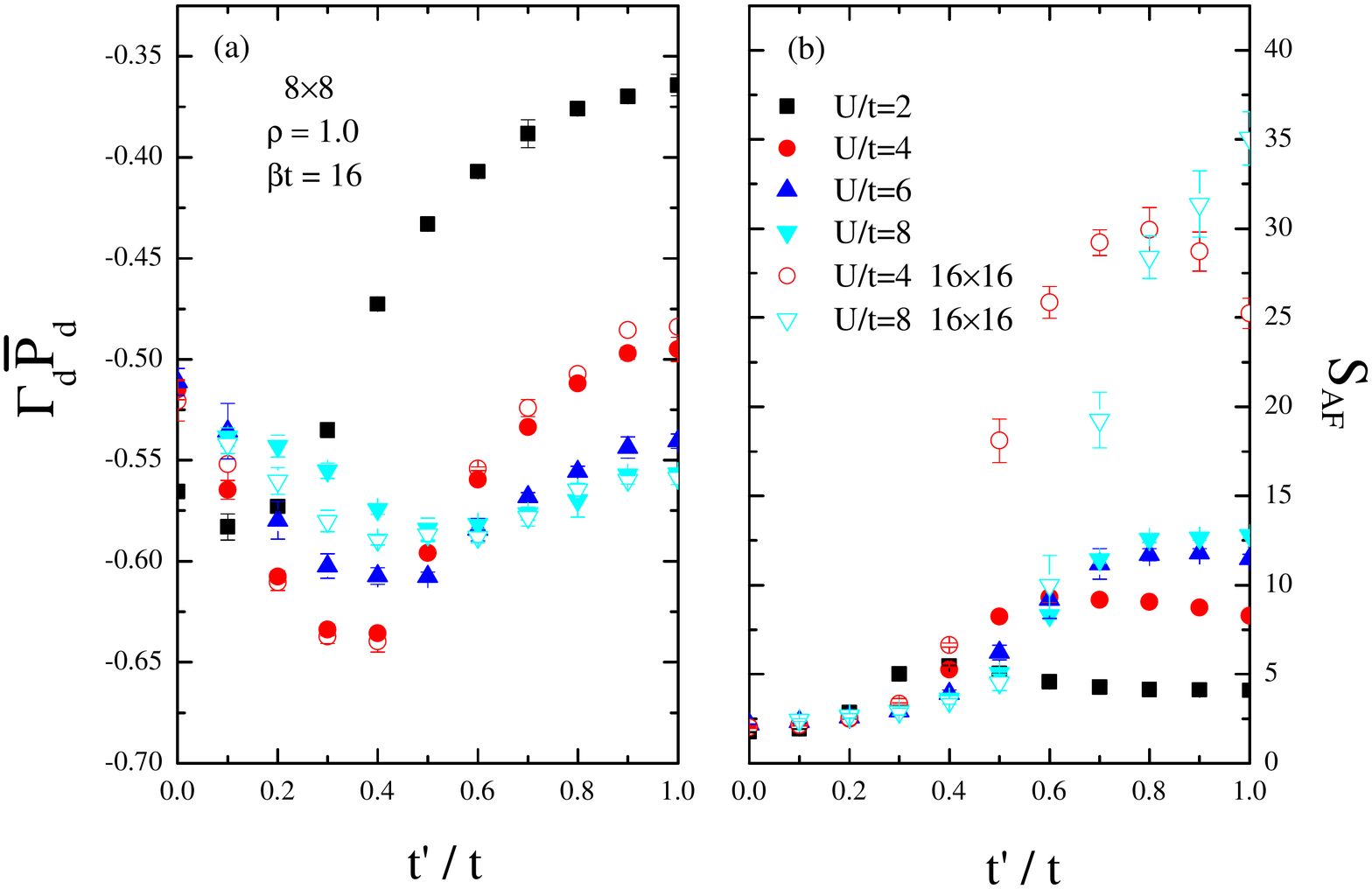,width=9.0cm,angle=-0,clip}
\caption{(Color online)
The tendency to (a) superconductivity $\Gamma \bar P_d$ and (b)
antiferromagnetic structure factor $S_{AF}$ as functions of
inhomogeneity $t'/t$ at fixed low temperature $\beta t=16$ and filling
$\rho=1$ for different $U/t$.
}
\label{fig:beta16mu0} 
\end{figure}

To understand this result better we show, in
Fig.\ref{fig:c_spin_real_space_comp}, the normalized difference of the
spin correlations,
\begin{eqnarray}
\delta c_{\rm spin}(\vec r \,) =
\frac{
 \, c_{\rm spin}^{t'=0.8}(\vec r\, ) - c_{\rm spin}^{t'=1.0}(\vec r\, ) \,
}
{ \, c_{\rm spin}^{t'=0.8}(\vec r\, ) + c_{\rm spin}^{t'=1.0}(\vec r\, ) \,
}
\end{eqnarray}
Panel (a)
has $\vec r$ along the (1,0) direction, and panel (b)
along the (1,1) direction.
With the
exception of the correlation for spins which are first neighbors in
different plaquettes, all values of $\vec r$ show an increase,
$\delta c_{\rm spin} > 0$.
What this tells us is that the enhancement in
$S_{\rm AF}$ comes from an increase in the real-space
spin correlations for
all separations, and is not simply from an enhancement at short
(or long) distances.  The fact that $\delta c_{\rm spin}(\vec r=0)$
is small further informs us that the effect is not just due to
a trivial change in the local moment.

We conclude this section by showing half-filled results for different
interaction strength $U/t$.  In Fig.~\ref{fig:U8mu0}(left panel), the
evolution with $t'/t$ of the product of the $d$-wave superconducting
vertex and the no-vertex susceptibility, $\Gamma_d \bar P_d$, is given
for $U/t=8$ and several temperatures.  As the temperature is lowered, a
clear minimum in $\Gamma \bar P_d$ indicates an optimal inhomogeneity,
at a larger $t'/t$ than for $U/t=4$.

In Fig.~\ref{fig:U8mu0}(right panel), the antiferromagnetic structure factor
is given.  As the temperature decreases, the antiferromagnetic
structure factor becomes larger. Similar to the case of $U/t=4$, $S_{AF}$
has an onset at $t'/t \sim 0.6$.

Figure \ref{fig:beta16mu0} shows
$\Gamma_d \overline P_d$ and $S_{AF}$
for four different $U/t$ values at fixed inverse temperature
$\beta t=16$. In Fig.~\ref{fig:beta16mu0}(a), an optimal inhomogeneity is
present for all $U/t$, shifting systematically to larger $t'/t$ as $U/t$
increases.
We note this trend is generally consistent with what is shown in
Fig.~4 of Ref.~[\onlinecite{fye92}],
Fig.~2 of Ref.~[\onlinecite{tsai08}],
and Figs.~2,5 of Ref.~[\onlinecite{baruch10}].
The maximum in $|\, \Gamma \bar P_d \,|$ is most evident at
$U/t=4$, for this fixed inverse temperature $\beta t=16$.  Comparison of
$S_{AF}$ data for 8x8 and 16x16 lattices shows that the structure factor
is growing roughly proportional to the volume, as expected in an ordered
Ne\'el phase.  The magnetic structure factor also increases with $U/t$
as double occupancy is suppressed.

\subsection{The doped lattice}

\begin{figure}[h]
\vspace{-0.2cm}
\epsfig{figure=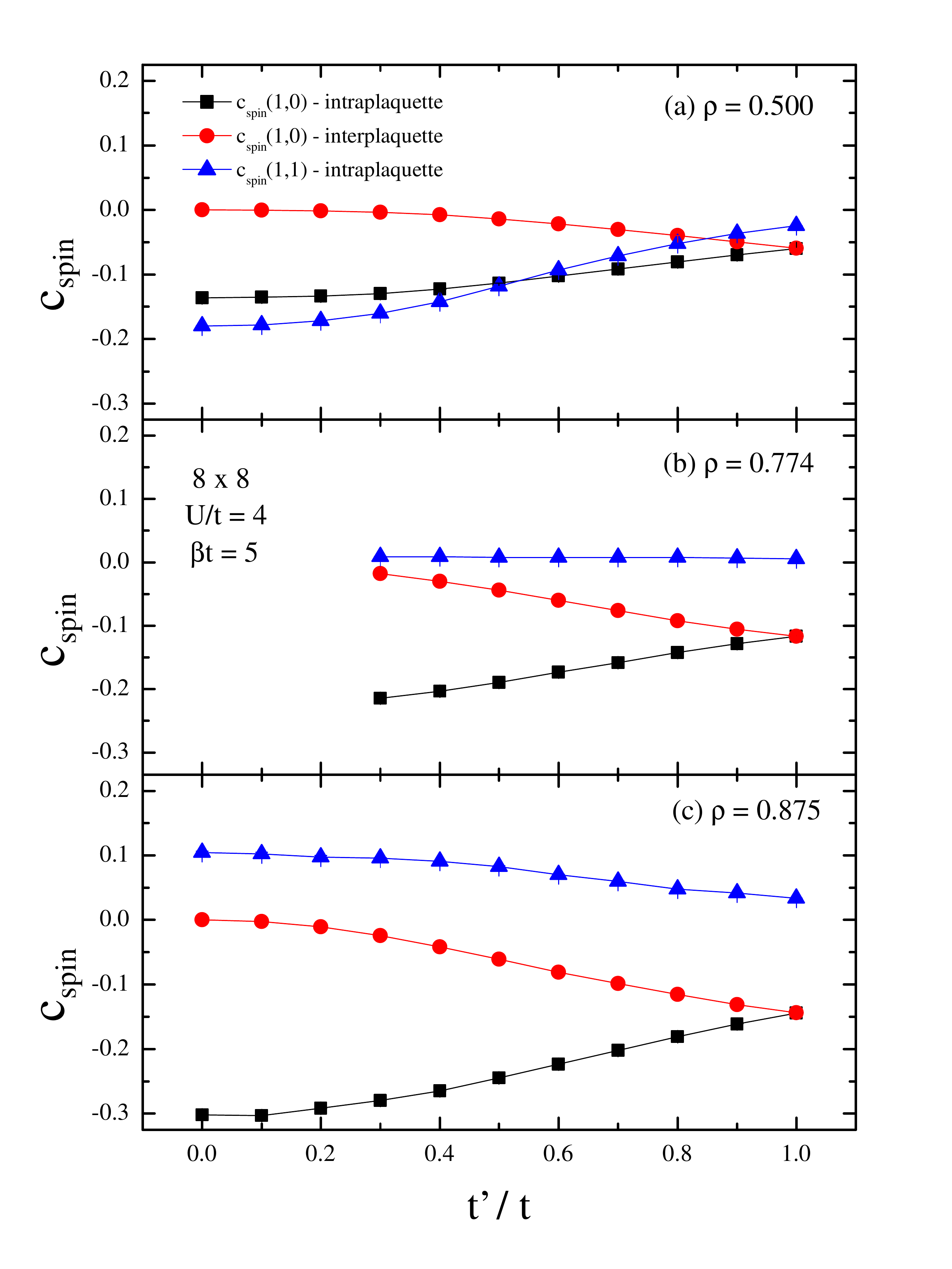,width=9.0cm,angle=-0,clip}
\vspace{-0.2cm}
\caption{(Color online)
Spin correlations $c_{\rm spin}(1,0)$ along an intraplaquette ($t$) bond
and along an interplaquette ($t'$) bond.  Also shown are next-near
neighbor values inside a plaquette.  The lattice size is 8x8, inverse
temperature $\beta t=5$ and interaction strength $U/t=4$.
\label{fig:Spincorrmerged-dopedL8}  
}
\end{figure}

After this brief synopsis of results at $\rho=1$, we turn to the case
when the filling is incommensurate, the situation of most interest to
understanding cuprate superconductivity.

Fig.~\ref{fig:Spincorrmerged-dopedL8} shows the same spin correlations
as in Fig.~\ref{fig:spinandcharge}(a), but for $\rho=0.500$ (a),
$\rho=0.774$ (b) and $\rho = 0.875$ (c).  The NN spin correlations
exhibit the expected evolution with density- they are largest at
$\rho=1.000$ (Fig.~\ref{fig:spinandcharge}(a)),  and decrease as we move
away from half-filling.  Similar to what happens at half-filling, the
NNN spin correlation inside a plaquette is positive for $\rho=0.875$,
again as expected for antiferromagnetism, but decreases with growing
$t'$.  With decreasing density the behavior for this quantity changes:
for $\rho=0.774$, $c_{spin}(1,1)$ is essentially zero for all $t'/t$.  For
$\rho=0.500$, however, it is \textit{negative} and increases in
magnitude as the connection between the plaquettes is reduced. This
later result can be understood when we recall that at this density the
system has two fermions in every plaquette on average. The configuration
which minimizes the kinetic energy and the local repulsive interaction
is a singlet state with spins residing on NNN neighbors. In this case
the NNN correlation becomes negative.  This effect is enhanced as $t'/t$
is smaller.

\begin{figure}[t]
\vspace{-0.8cm}
\epsfig{figure=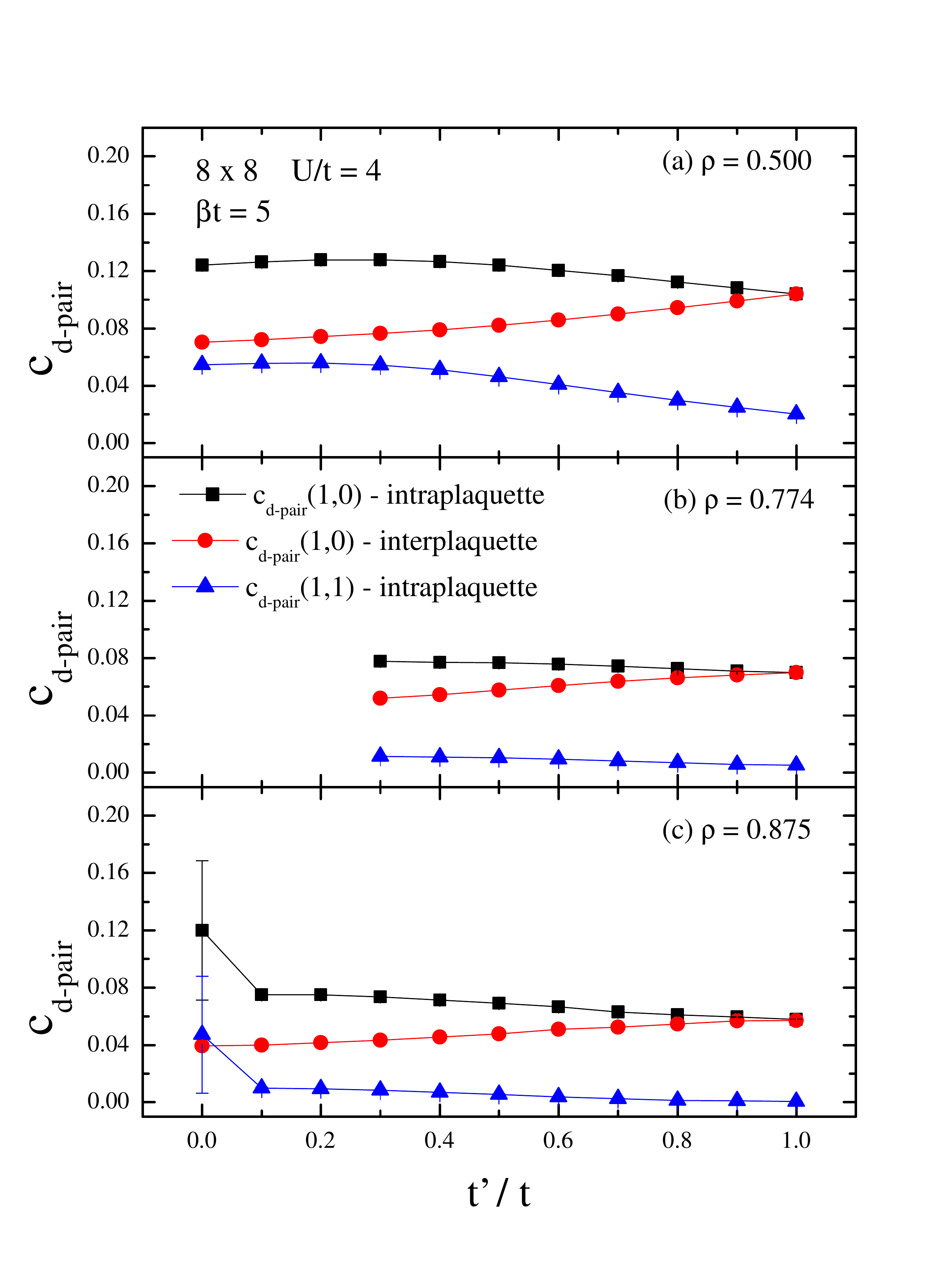,width=9.5cm,angle=-0,clip}
\vspace{-0.2cm}
\caption{(Color online)
Near-neighbor and next-near neighbor
$d$-wave pairing correlations as a function of $t'/t$.
The lattice size is 8x8, inverse temperature $\beta t=5$ and
interaction strength $U/t=4$.
\label{fig:C-dpairmerged-dopedL8}  
}
\end{figure}

Figure~\ref{fig:C-dpairmerged-dopedL8} shows  short-range $d$-wave pair
correlations for the same densities as
Fig.~\ref{fig:Spincorrmerged-dopedL8}.  Contrary to what is observed for
spin correlations,  NN and NNN pairing correlations do not decrease with
doping.

\begin{figure}[h]
\vspace{-0.2cm}
\epsfig{figure=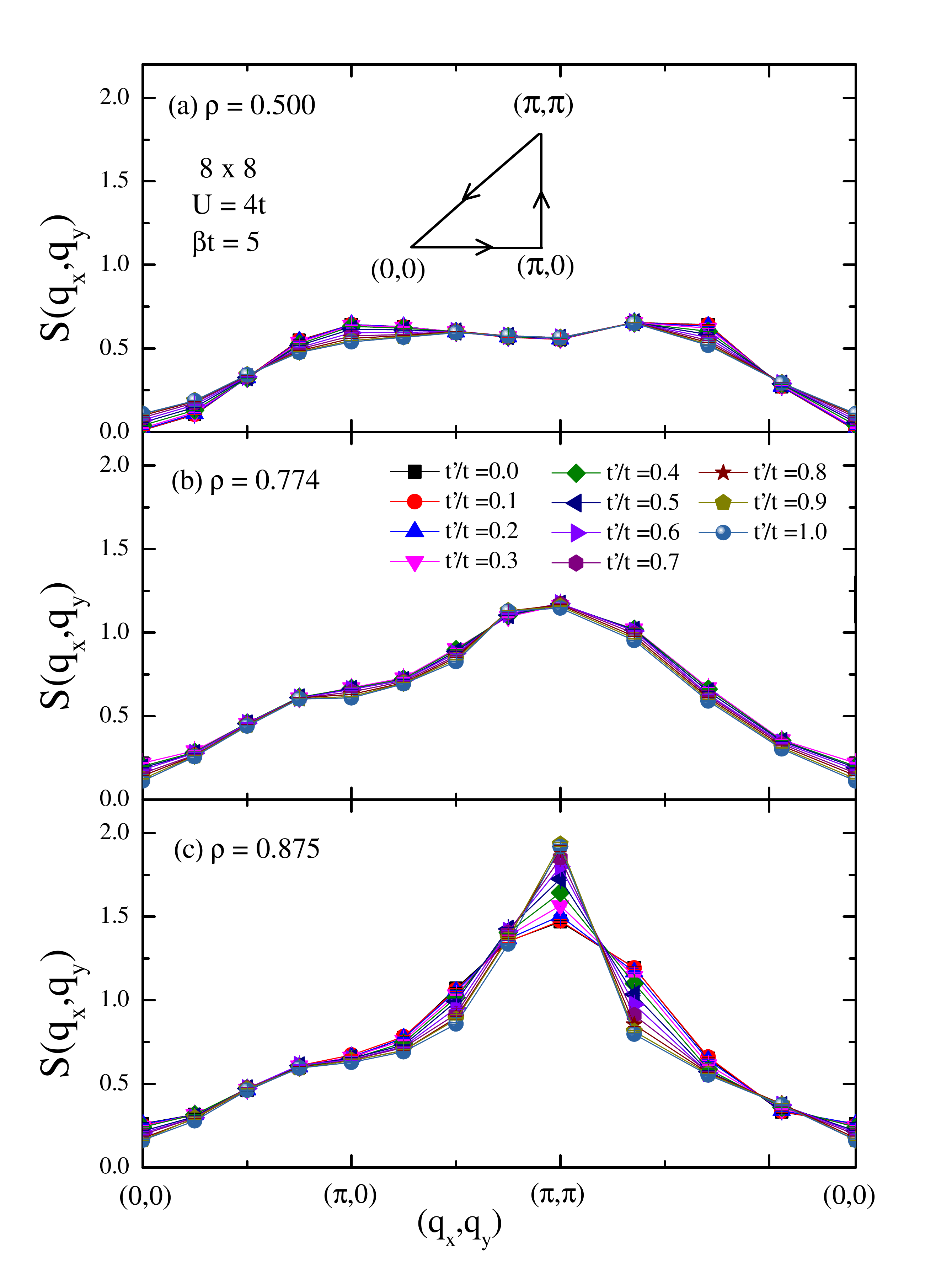,width=9.0cm,angle=-0,clip}
\vspace{-0.2cm}
\caption{(Color online)
The equal time structure factor $S(q_x,q_y)$ is shown as a function of
momentum as one traverses the Brillouin zone triangle shown in the panel
(a) inset.  Panels (a), (b)  and (c) correspond to the dopings
$\rho=0.5, 0.774,$ and $0.875$ respectively.  For each density, $S$ is
given for eleven different $t'/t$.  The lattice size is 8x8, $U/t=4$,
and inverse temperature $\beta t = 5$.  \label{fig:Sqxqybeta5} 
}
\end{figure}

Having described the short range, real space correlations, we now turn
to more sensitive magnetic and pairing structure factors and
susceptibilities.  The latter especially has an enhanced signal since it
is sensitive to the build-up of correlations in the imaginary time
direction.  The magnetic structure factor dependence  on $t'/t$ and
$q_x,q_y$ for three different dopings on an 8x8 lattice at inverse
temperature $\beta t=5$ is shown in Fig.~\ref{fig:Sqxqybeta5}.  Near
half-filling ($\rho=0.875, 0.774$) $S(q_x,q_y)$ is peaked at
$(\pi,\pi)$, indicating the dominance of antiferromagnetic correlations.
At $\rho=0.875$ the AF peak substantially increases as $t' \rightarrow
t$, with a concomitant reduction in $S$ at other momenta.  Presumably
these effects would become larger at lower $T$.  However, $\beta t
\approx 5$ is the limit accessible to DQMC owing to the sign problem.
For lower densities, $S(q_x,q_y)$ is rather insensitive to $t'$.

There is a substantial difference in scale of the antiferromagnetic
structure factor: $S_{\rm AF}=S(\pi, \pi) \sim 1$ in the doped lattice,
whereas at half-filling, $S_{\rm AF} \sim 10$ (Fig.~\ref{fig:U4mu0}).
This arises both from the rapid suppression of antiferromagnetic order
with doping in the square lattice Hubbard model\cite{hirsch85,white89a},
and also because of the lower temperatures that can be reached at
$\rho=1$ ($\beta t\sim 10-16$) compared to $\rho \neq 1$ ($\beta t \sim
5$).  Despite the absence of long range order in the doped model,
the short range spin correlations do grow as $T$ is lowered.

At $\rho=0.875$ the overall evolution with $t'/t$ of the antiferromagnetic
structure factor $S(\pi,\pi)$ in Fig.~\ref{fig:Sqxqybeta5} is consistent
with that found in [\onlinecite{baruch10}].  That is, $S(\pi,\pi)$
increases monotonically with $t'/t$ and is maximal at $t'/t=1$. However,
the two results appear to differ in the finer details.  Specifically,
the CORE study indicates that the staggered magnetic order parameter is
roughly constant for $ 0 < t'/t <0.5$, and then increases rather
abruptly at $t'/t \approx 0.6.$ This is mirrored in an increase in the
number of magnons, a phenomenon to which the appearance of a maximum in
the pair binding energy is attributed.  In contrast, our DQMC data
appear to indicate a more immediate rise in $S(\pi,\pi)$ as $t'$ grows
from zero.  A possible origin of the difference is that our work is at
finite temperature, whereas the CORE study is in the ground state.
Indeed, at half-filling it is known that $S(\pi,\pi)$ does not reach its
low $T$ values until $T/t \lesssim 0.08$, temperatures which are not
accessible when the system is doped, due to the sign problem.  That
finite temperature is a likely explanation of the difference and is
substantiated by examining the $\rho=1$ data in Fig.~\ref{fig:U4mu0}.
Interestingly, the rapid rise in $S_{\rm AF}$ occurs at the same $t'/t
\sim 0.6$ obtained from CORE.  Note that in [\onlinecite{baruch10}] the
number of holes is $N_h=2,4$ on a 6x6 cluster, corresponding to $\rho =
0.945, 0.890$.  The latter value is comparable to that of
Fig.~\ref{fig:Sqxqybeta5}(c).

\begin{figure}[t]
\epsfig{figure=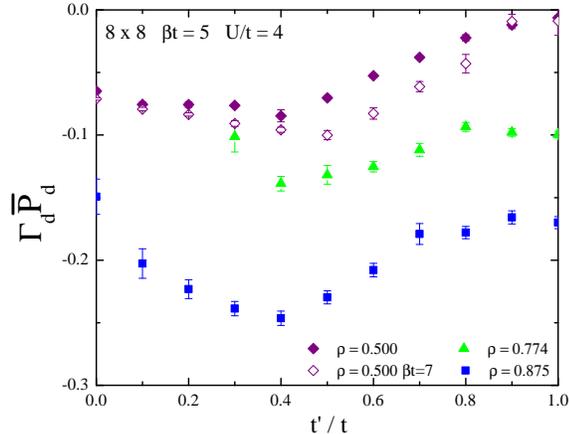,width=9.0cm,angle=-0,clip}
\caption{(Color online)
Dependence of the pairing vertex on $t'/t$ away from half filling.
All densities appear to show a maximum of $|\Gamma_d \overline P_d|$
at intermediate $t'/t$.
The lattice size is 8x8, $U/t=4$ and $\beta t = 5$. Also
included for comparison is the $\beta t = 7$ data in the $\rho=0.500$ case.
\label{fig:GammaPdnvvstbeta5U48x8}  
}
\end{figure}

Figure \ref{fig:GammaPdnvvstbeta5U48x8} extends the pairing results of
Fig.~\ref{fig:U4mu0}(a) to the doped case. Although
the sign problem currently prevents simulations at low $T$,
all densities shown exhibit a
maximum in $|\Gamma_d \overline P_d|$ away from the uniform limit
$t'/t=1$.  For $\rho=0.774$ and $\rho=0.875$, the sign is fairly small
for $\beta t=5$ and small $t'/t$ (see Fig. 2).  We have thus done a
very large number of runs (up to 100 runs with 50000 sweeps each), to
decrease the error bars and push the limits of the QMC method.  For
$\rho=0.5$, where the sign is higher, we were able to reach $\beta t=7$ (open symbols).  Similar to what is seen in Fig.~\ref{fig:U4mu0}(a),
for $\rho=1$, the signal for optimal inhomogeneity  indeed increases
as $T$ is lowered. It is reasonable to assume that the same trend will
hold for $\rho=0.774$ and $\rho=0.875$.

Almost all of the work presented in this paper is for $U/t=4$.  The sign
problem in DQMC becomes dramatically worse as $U/t$ increases.  In order
to study the $U/t$ evolution and still reach reasonably low
temperatures, we can reduce the density to $\rho=0.5$ which restores the
sign even though $U/t \gtrsim 4$.
Even so, it is not possible for us to assess accurately
claims\cite{tsai08,baruch10,karakonstantakis11} that $U/t \sim 8$ is
optimal for pairing.

\begin{figure}[t]
\epsfig{figure=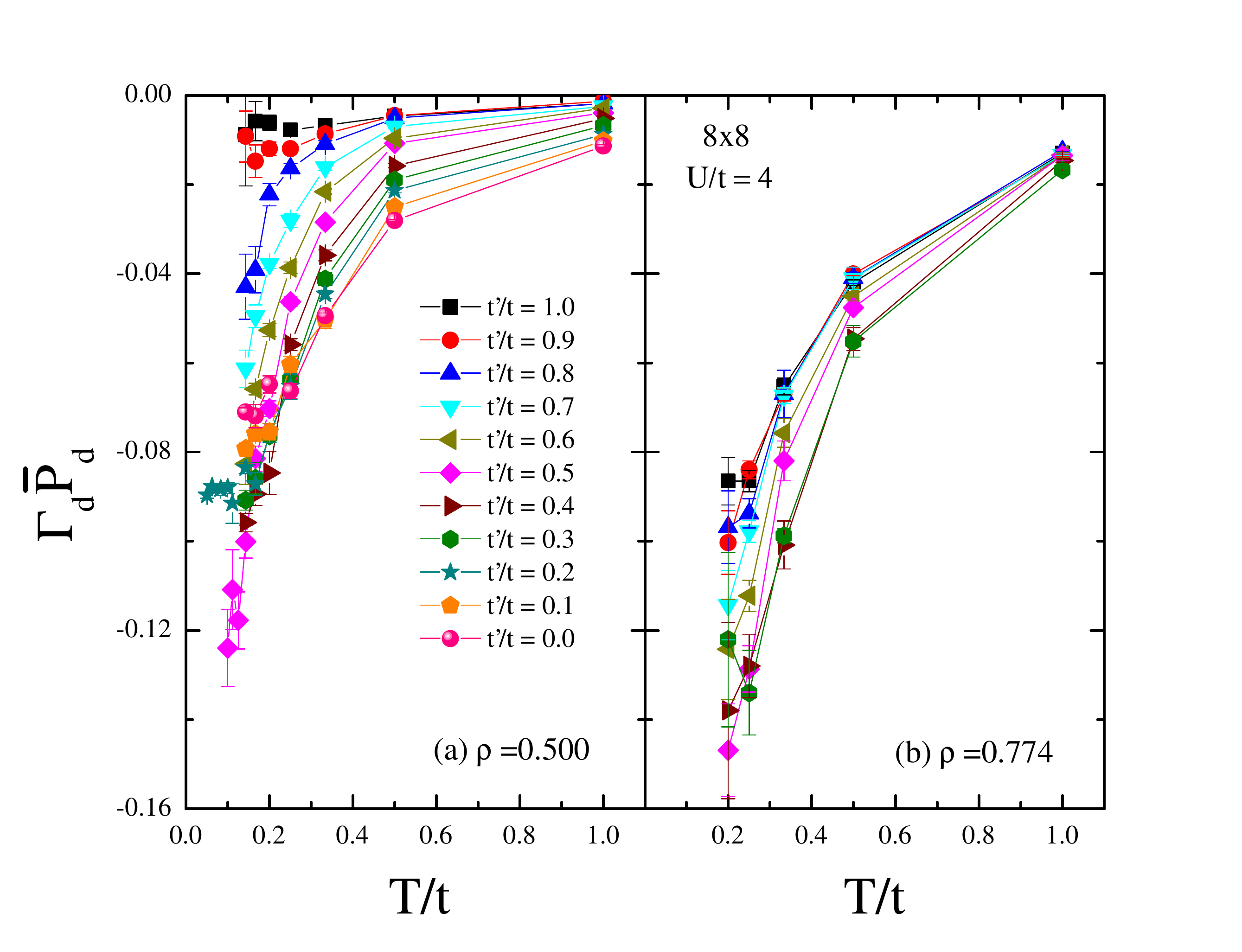,width=9.5cm,angle=-0,clip}
\caption{(Color online)
The evolution of $\Gamma_d \overline P_d$ with $T/t$ for  $\rho=0.500$ (a) and
$\rho=0.774$ (b).
The plots emphasize the existence of an optimal degree of inhomogeneity.
Here the lattice size is 8x8 and $U/t=4$.
\label{fig:GammaPdnvvsT}  
}
\end{figure}

The choice $\rho=0.5$ does however improve the average sign enough to
see the $t'/t$ evolution of $d$-wave pairing, which we established to
have an optimal inhomogeneity at half-filling.
Fig.~\ref{fig:GammaPdnvvsT} (a) shows $\Gamma_d \overline P_d$ versus
$T/t$ for different $t'/t$.  As at $\rho=1$, there is evidence for an
optimal inhomogeneity:  in the uniform case $\Gamma_d \overline P_d$
versus $T/t$ is almost temperature independent and is also small,
$|\Gamma_d \overline P_d| \lesssim 0.01$.  As inhomogeneity is turned on
to $t'/t \sim 0.5$, $|\Gamma_d \overline P_d|$ increases by almost an
order of magnitude (although it is still far from the $\Gamma_d
\overline P_d = -1$ criterion for a transition).  Further increase of the
inhomogeneity to $t'/t < 0.5$ decreases $|\Gamma_d \overline P_d|$.  The
same optimum $t'/t \sim 0.5$ can be seen for $\rho=0.774$, as shown in
Fig.~\ref{fig:GammaPdnvvsT}(b).

\begin{figure}[h]
\epsfig{figure=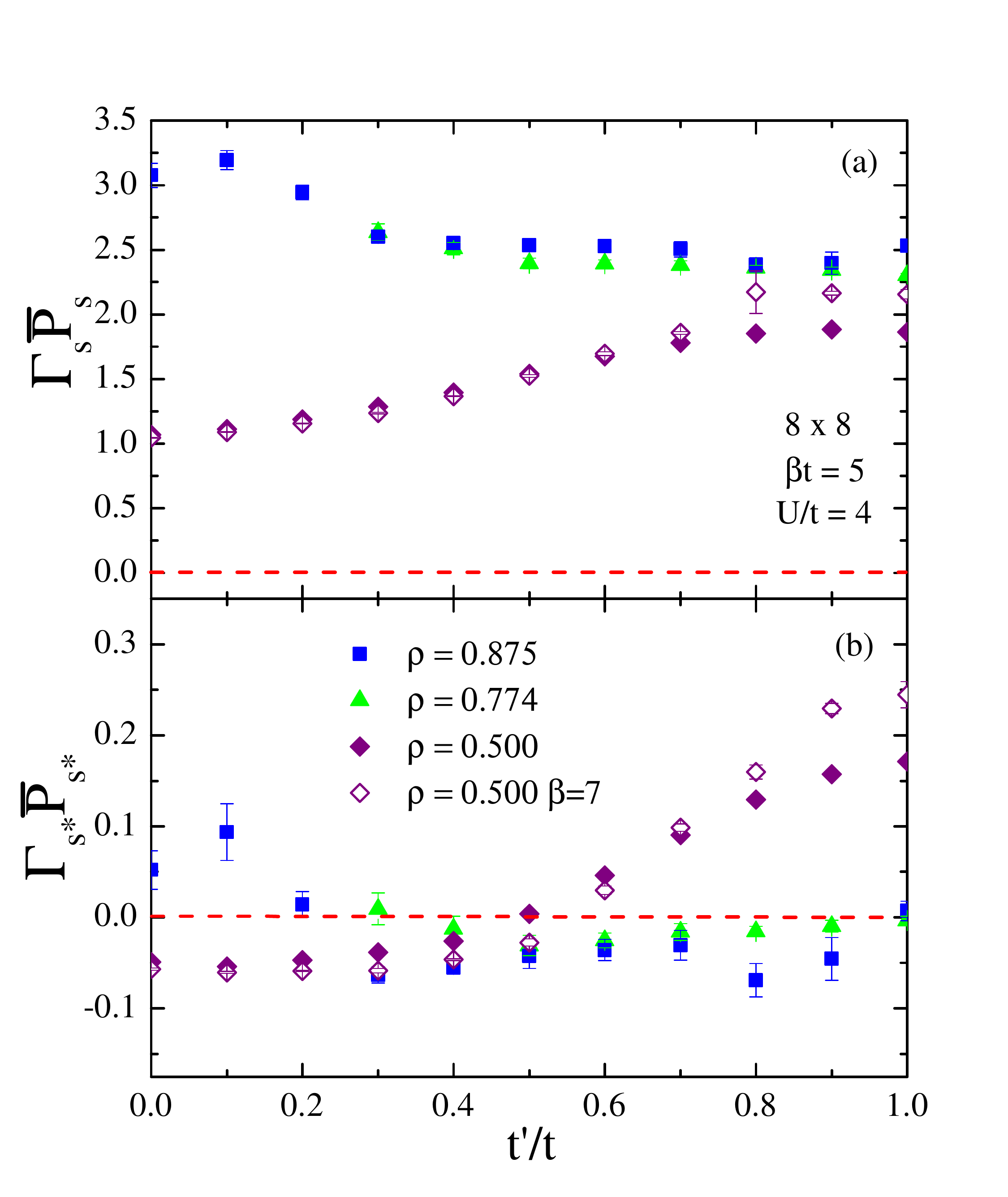,width=9.0cm,angle=-0,clip}
\caption{(Color online)
Same as Fig.~\ref{fig:GammaPdnvvstbeta5U48x8} but now comparing the interaction
vertex
times the uncorrelated susceptibility for two other symmetry channels: $s$ in
(a) and $s^*$ in (b). While in the former all densities result in a repulsion
between the pairs for the whole range of $t'/t$ studied, in the latter
depending on the specific parameters the pairing turns attractive but
is substantially
smaller in magnitude in comparison to the $d$-wave symmetry channel.
\label{fig:GammaPsandPsx}  
}
\end{figure}

Early in DQMC studies of the homogeneous square lattice it
was established that $d$-wave pairing is the dominant
superconducting instability.  This conclusion is not
altered by $t' \neq t$.
Fig.~\ref{fig:GammaPsandPsx} shows  the same quantities as
Fig.~\ref{fig:GammaPdnvvstbeta5U48x8} for $s$ and
\textit{extended} $s$ ($s^*$) symmetry channels. The correlations are obtained
in a similar fashion as in the $d$-wave case but the associated phases in
Eq.~\ref{eq:pairsusc} are positive. For the $s$ symmetry the pairs are
created and destroyed locally $\left(  \Delta^{\dagger}_{s\,\vec r} =
c^\dagger_{\vec r\uparrow} c^{\dagger}_{\vec r\downarrow}\right)$, whereas  in
the
extended one they all enter with the same phase sign
$\left(\Delta^{\dagger}_{s^*\,\vec r} =
c^\dagger_{\vec r\uparrow} (c^{\dagger}_{\vec r+\hat x\downarrow}
+c^{\dagger}_{\vec r+\hat y\downarrow}
+c^{\dagger}_{\vec r-\hat x\downarrow}
+c^{\dagger}_{\vec r-\hat y\downarrow} )\right)$. While $s$-wave
symmetry produces
only repulsive interactions, some parameters in the $s^*$-wave case
exhibit attraction.  Nonetheless it is smaller in magnitude than
$d$-wave symmetry.

\section{Checkerboard Hubbard Model}

\begin{figure}[t]
\epsfig{figure=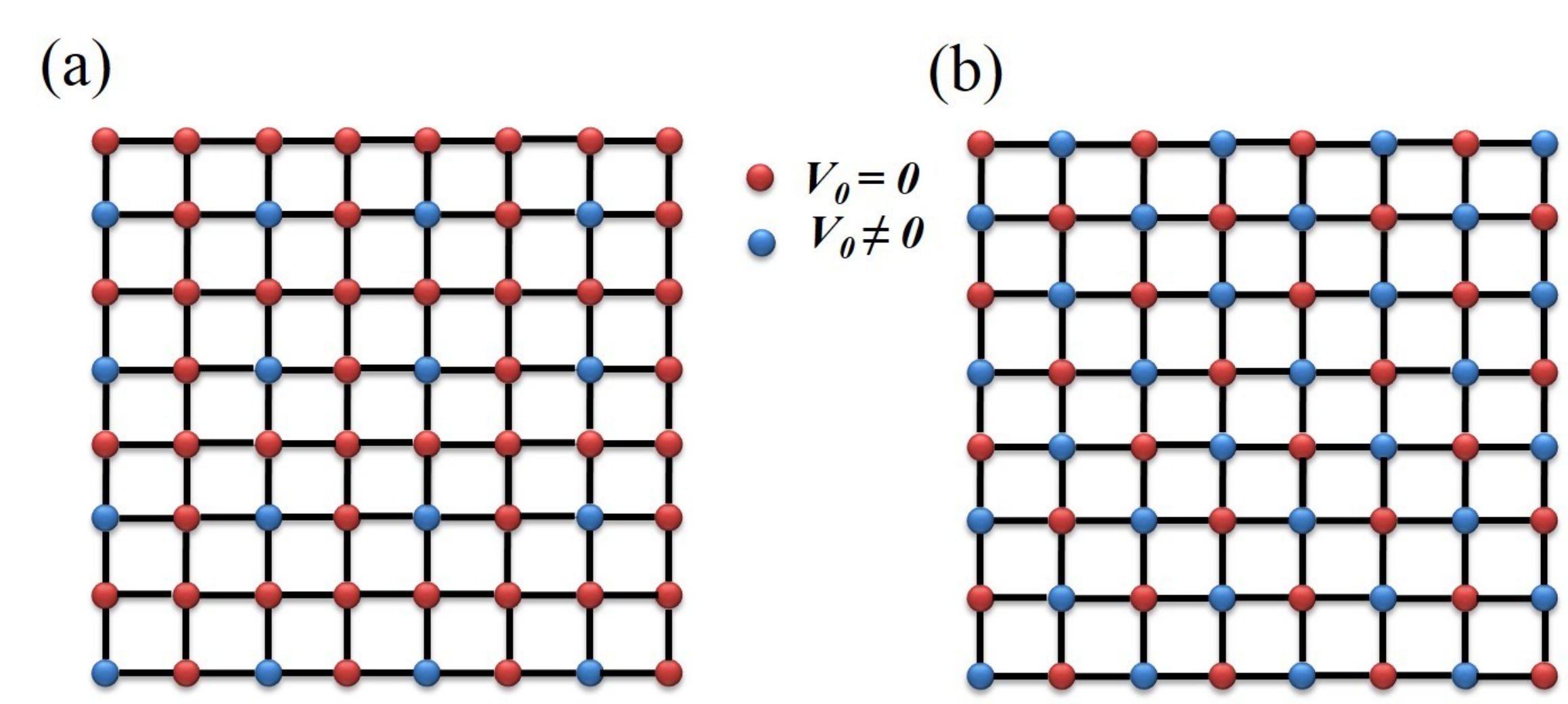,width=3.0in,angle=-0,clip}
\caption{(Color online)
Checkerboard geometry, in which a fraction $f$ of the sites, displayed in a
checkerboard pattern,  has  on site energy raised by $V_0 \ne 0 $.  Panel (a)
shows the  $f=1/4$ lattice and and (b) the $f=1/2$ one, for 8x8 systems.
\label{fig:checker}
}
\end{figure}  

\begin{figure}[h]
\epsfig{figure=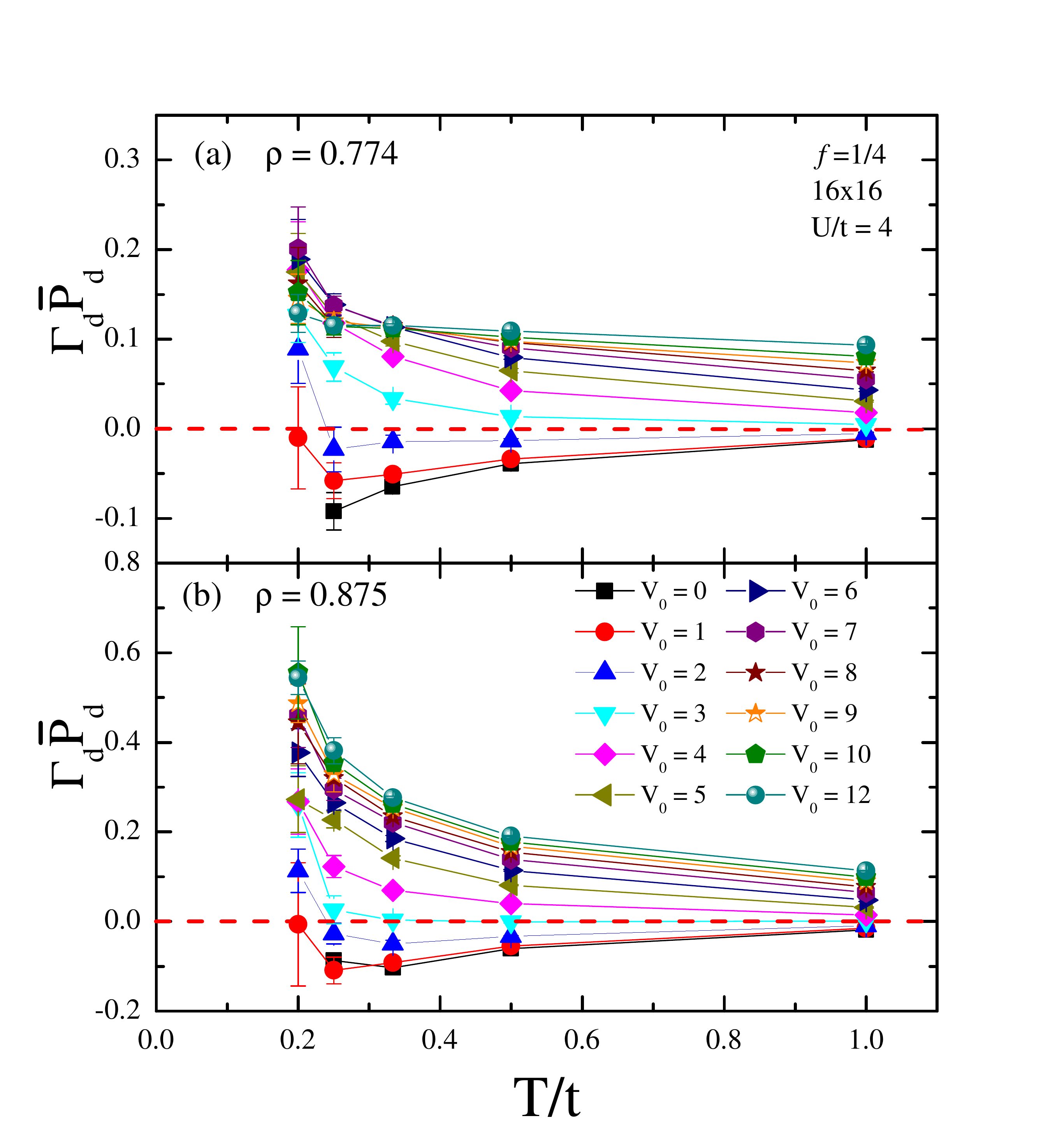,width=9.0cm,angle=-0,clip}
\caption{(Color online)
Product of interaction vertex $\Gamma_d$ and uncorrelated susceptibility
$\overline P_d$ for a Hubbard model with an alternating pattern of
site energies.  (See Fig.~\ref{fig:checker}(a).)
The vertex is weakly attractive for the homogeneous case, $V_0=0$,
but becomes repulsive for $V_0 \gtrsim 1$.  Here the lattice size is $16$x$16$,
filling $\rho=0.774$ in $(a)$ and $\rho=0.875$ in $(b)$, and interaction
strength $U/t=4$.
\label{fig:GammaPdnvrho0774and0875P4}
}
\end{figure}  

The nature of pairing in models with other sorts of inhomogeneities,
e.g.~built of two site dimers rather than four site
clusters,\cite{tsai06} modulated by different site
potentials\cite{Okamoto10} or consisting of lines of different chemical
potentials, alternating between half-filled antiferromagnetic stripes and
doped stripes has also been explored.\cite{mondaini12b}  In this section we
examine the effects on pairing of an inhomogeneity pattern in which
the local energies on a regular pattern of sites
is raised by an amount $V_0$.  That is, we add a term
$H'= V_0 \sum_{l \in {\cal A}, \sigma} n_{l\sigma}$
to the Hubbard Hamiltonian
Eq.~\ref{eq:plaquetteham}
with $t'=t$.
The collection ${\cal A}$ consists of  a fraction $f$  of the lattice sites.
This geometry is illustrated in Fig.~\ref{fig:checker}, for $f=1/4$ (a) and
$f=1/2$ (b), the two cases
analyzed here.

\begin{figure}[h]
\epsfig{figure=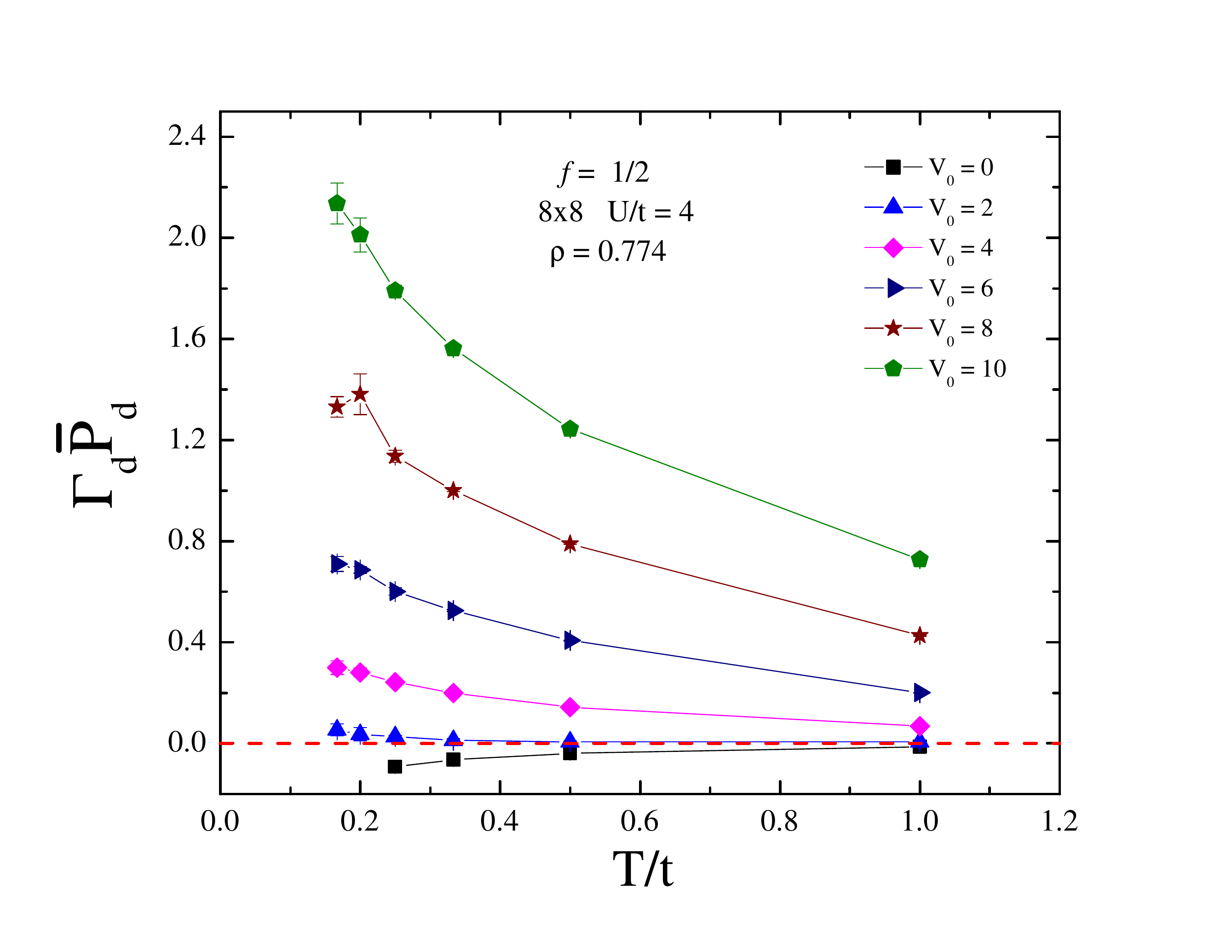,width=9.0cm,angle=-0,clip}
\caption{(Color online)
Same  as Fig.~\ref{fig:GammaPdnvrho0774and0875P4}, but for $f=1/2$ and
an 8x8 lattice.  The qualitative behavior is similar to the $f=1/4$
case:  The site inhomogeneity drives the interaction vertex repulsive.
\label{fig:GammaPdnvrho0774P2} 
}
\end{figure}

One motivation for considering this particular pattern with $f=1/4$  is
that in the limit $V_0 \rightarrow \infty$ the lattice maps onto the
`three band' Hamiltonian sometimes used to model the CuO$_2$ plane of
the cuprate superconductors (with, however, the choice of equal copper
$d$ and oxygen $p$ energies.) The red sites without any blue neighbors
are like the Cu atoms, while the red sites with two blue neighbors
represent the O sites which link the Cu.  Thus this model makes partial
contact with earlier studies of binding on CuO$_2$ clusters in the limit
$\epsilon_{pd}=0$.\cite{hirsch88a,balseiro88,hirsch89a} Another point of
contact of this model is to other inhomogeneity patterns which share an
$f=1/4$ proportion of sites with raised on-site energy, for example
[\onlinecite{mondaini12b}] in which a pattern of stripes was shown to
enhance $d$-wave pairing away from half-filling.

Results for this site-energy inhomogeneous geometry ($f=1/4$) are shown
in Fig.~\ref{fig:GammaPdnvrho0774and0875P4}(a).  In stark contrast to
the plaquette model and to the \textit{striped} $V_0$
model,\cite{mondaini12b} $\Gamma_d \overline P_d$ becomes positive when
$V_0$ is turned on:  the $d$-wave pairing vertex is made repulsive.  As
with the plaquette Hamiltonian of the previous sections, we are
interested in how the dependence of pairing on inhomogeneity is affected
by the density.  To  this end we show, in
Fig.~\ref{fig:GammaPdnvrho0774and0875P4}(b), the same quantity but with
$\rho=0.875$.  This data is consistent with the previous density, and we
conclude that this form of site energy inhomogeneity competes
destructively with superconductivity.

Finally, we consider a pattern of inhomogeneity with $f=1/2$.
(see Fig.~\ref{fig:checker}(b).)
Fig.~\ref{fig:GammaPdnvrho0774P2}.
demonstrates the effect of $V_0$ is monotonic and again inimical
to superconductivity.

\section{The Sign Problem}

We return here briefly to the sign problem which is the fundamental
limitation to using DQMC to study lattice fermion Hamiltonians
\cite{loh90}.
In generic situations (that is, in the absence of particle-hole
or some other symmetry which makes the sign positive)
the fermion sign
$\langle {\cal S} \rangle$
is well-behaved down to some temperature scale
$T/t \sim \alpha$, where
$\frac13 \lesssim \alpha \lesssim \frac15 $ depends, in the single band
Hubbard model, on $U$ and $\rho$.
Below this temperature, $\langle {\cal S} \rangle$
decays exponentially with $\beta=1/T$ so that simulations
are feasible only in a very narrow range of temperatures
below the point at which some of the fermion determinants begin to go negative.
$\langle {\cal S} \rangle$ also decays exponentially with
spatial volume $V$,
but in practice the $\beta$ dependence is
usually more problematic.
The behavior of
$\langle {\cal S} \rangle$ with $\rho$ and spatial geometry
is also affected by `shell effects',\cite{mondaini12a} so that
the sign can remain close to unity for fillings for which
multiple $k$ points have the same non-interacting
energy $\epsilon_k$.

\begin{figure}[t]
\epsfig{figure=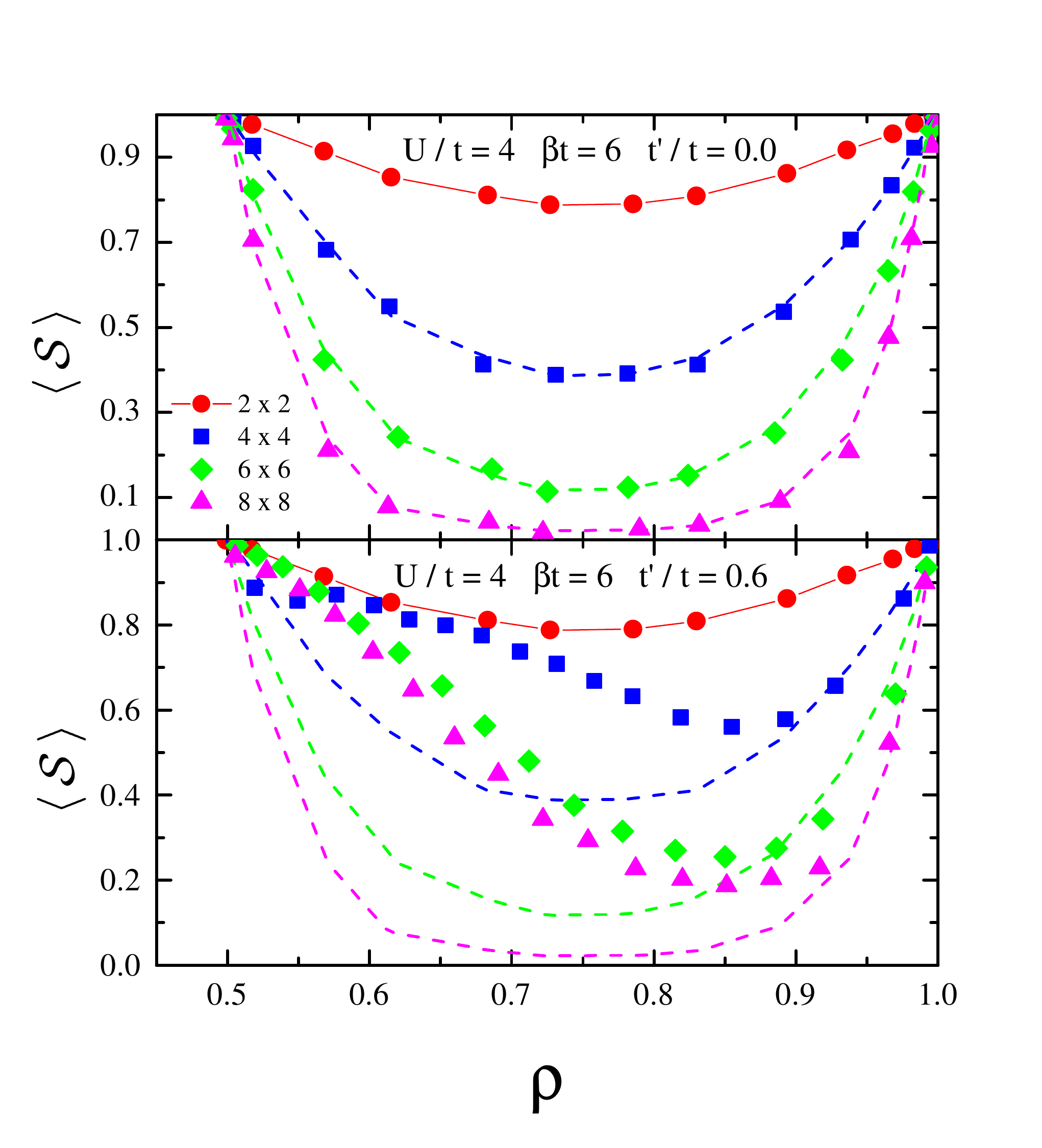,width=9.0cm,angle=-0,clip}
\caption{(Color online)
\underbar{Top panel:}
Average sign
$\langle {\cal S} \rangle$
as a function of filling $\rho$ for $t=1, U/t=4, \beta t=6$
and different system sizes.
Interplaquette hopping $t'/t=0$ so that the system
is composed of $(L/2)^2$ independent $2 \times 2$ clusters.
Dashed lines are the prediction for the average sign obtained
by taking the sign for a $2 \times 2$ lattice
(i.e.~a single plaquette) raised to the $(L/2)^2$ power.
\underbar{Bottom panel:}  Same except interplaquette hopping $t'/t=0.6$.
The fact that
$\langle {\cal S} \rangle \sim 1$
for $\rho \sim 0.5$ is a `shell effect'.  (See text.)
\label{fig:sign} 
}
\end{figure}

The plaquette Hubbard Hamiltonian offers a window into
this $V$ dependence, since it must be rigorously true
at $t'/t=0$ that
$\langle {\cal S}(V=L\times L) \rangle
= \langle {\cal S}(V=2\times 2) \rangle^{(L/2)^2}$.
It is interesting, then, to understand how the coupling
of independent plaquettes with $t'/t \neq 0$ modifies this
manifestly exponential decay.
We show results in Fig.~\ref{fig:sign}.
When $t'/t=0$ (top panel) the average sign (symbols)
precisely follows the prediction (dashed lines)
based on the
sign of an elemental $2 \times 2$ cluster.
However, when the clusters are coupled, $t'/t=0.6$
(bottom panel),
the average sign is increased.
While the improvement in the behavior of
$\langle {\cal S} \rangle$
is not sufficient to allow ground state properties
to be obtained, it is nevertheless intriguing,
and non-trivial, that the
entanglement of the clusters by hopping $t'$
here reduces the sign problem:
The coefficient $\gamma$ of the exponential decay
$\langle {\cal S} \rangle \propto e^{-\gamma V}$
changes from $\gamma \sim 0.056$, at $t'/t=0.0$, to
$\gamma \sim 0.014$, at $t'/t=0.6$.

\section{Conclusions}

Study of the effect of inhomogeneities on superconductivity has been a
focus of much computational effort on the Hubbard and $t-J$ models over
the last decade.  One branch of effort has explored models where
inhomogeneity is included in the Hamiltonian itself.  Other work
concerns the question of inhomogeneity which arises spontaneously in a
translationally invariant Hamiltonian.  The plaquette Hubbard model has
been a natural candidate of interest since it seems to contain the
nascent element, a substantial binding energy, in its building blocks.

We have shown here that DQMC indicates that a sensitive
measurement of $d$-wave pairing yields an `optimal degree of
inhomogeneity'.  That is $\Gamma \overline P_d$ is closer to $-1$ at
$t'/t \sim 0.4$ when $U/t=4$
than at $t'/t =0$ or $t'/t =1$.
Larger $U/t$ lead to larger optimal $t'/t$ for the
$2 \leq U/t \leq 8 $ range studied.
This result agrees
qualitatively with some past numeric work (differing in the precise
optimal $t'/t$), but is in disagreement with several of the most
powerful computational methods available for these sorts of problems.
When simulations are conducted directly on doped lattices, our work
clearly shows the existence of an optimal inhomogeneity, which develops
further as $T$ is lowered.  While the sign problem prevents us from
going to very low temperatures, we can further infer what happens for
small doping through our results at half-filling, where there is little
limitation on the accessible temperature.  Here, as White {\it et.al.}
have emphasized\cite{white89a} the pair correlation function $ c_{d
\,{\rm pair}}({\vec r,\tau}\,)= \langle \Delta^{\phantom{\dagger}}_{d\,
\vec r_0 + \vec r}(\tau) \, \Delta^{\dagger}_{d\,\vec r_0}(0) \rangle $
probes the insertion and propagation of a pair of fermions in the
half-filled lattice, resulting in an effective doping $\delta = 2/L^2$.
On an 8x8 lattice, for example, this corresponds to $\delta \sim 0.03$.
Direct simulations of the doped system would, of course, be preferable,
especially since this effective doping is system size dependent.
Nevertheless, comparisons of data for different $t'/t, U/t$ and (large)
$\beta t$ on lattices of {\it fixed} size are possible.  In this way, our
low temperature, half-filled results speak to the issue of the role of
inhomogeneity.
Simulations of an alternative `checkerboard Hubbard model' show that the
attractive d-wave vertex is not generic.  A different pattern of spatial
inhomogeneity produced a repulsive vertex for most parameter regimes of this
second Hamiltonian.

We have also exploited the ability to decouple the model spatially
to obtain data on the sign problem as independent ($t'/t=0$) spatial
clusters are coupled.  We find that the average sign
is increased by finite $t'/t$.

We conclude this paper with a few remarks concerning the results from
different numerical methods.  Our DQMC results are most consistent with
the exact diagonalization\cite{tsai08}, DMRG\cite{karakonstantakis11},
and CORE\cite{baruch10} treatments, indicating the existence of an
optimal degree of inhomogeneity.  As is well known, what makes the
problem of strong correlation so challenging is the competition between
different possible ordered (or disordered) states at low temperature
which are close in energy.  Even small approximations in analytic and
numeric treatments can tip the balance in these near-degeneracies.  DQMC
treats the correlated electron problem exactly on lattices of finite
size.  Here, we are exploring a superconducting mechanism which
explicitly attributes pairing to a spatially local attraction, as
opposed to the exchange of lattice vibrations or spin waves.  In such a
situation it is plausible that finite size effects, while still present,
might be less strong.


\vskip0.15in
{\bf Acknowledgements:}
This work was supported by the National Key Basic Research Program of
China, Grant No.~2013CB328702, by DOE DE-NA0001842-0, and by the Office
of the President of the University of California.  Support from  CNPq
and FAPERJ (TP and RM) is gratefully acknowledged.



\begin{thebibliography}{100}
\bibitem{foot1} It is worth noting that there is a considerable
literature on another model sometimes also referred to as the
``checkerboard Hubbard Hamiltonian" (or ``2D pyrochlore"
Hamiltonian).\cite{yoshioka08,trousselet08,poilblanc07,huang07,indergand07,Raczkowski09}
In this model, hopping is present along the diagonal of alternating
plaquettes of the square lattice Hubbard Hamiltonian.  One key question
is whether a spin liquid phase in which there is no symmetry breaking
emerges due to the frustration, or whether the ground state retains some
sort of translation-invariance-breaking or exotic pairing order.  The
model considered in [\onlinecite{yao10}]
contains elements both of a Hamiltonian built up of linked
plaquettes and the frustrating effects of diagonal hopping.
\end{thebibliography}
\end{document}